%% file: phase_diagram_soft_disks.tex
\documentclass[notitlepage,english,aps,prx,tightenlines,floatfix,10pt,longbibliography,twocolumn,superscriptaddress,footinbib]{revtex4-2}

\usepackage{graphicx}
\usepackage{dcolumn}
\usepackage{bm}
\usepackage[dvipsnames]{xcolor}
\usepackage{amssymb,ulem,amsmath}
\usepackage{physics}
\usepackage{layouts}
\usepackage[colorlinks=true,urlcolor=blue,citecolor=blue,linkcolor=blue]{hyperref}
\usepackage{orcidlink}

\newcommand{\qql}{\textquotedblleft}
\newcommand{\qqr}{\textquotedblright}

\newcommand{\dg}{\dagger}

\newcommand{\ham}{\hat{\mathcal{H}}}

\newcommand{\fpsi}{\widetilde{\psi}}

\begin{document}
	
\author{S. Peotta\orcidlink{0000-0002-9947-1261}}
\affiliation{Department of Applied Physics, Aalto University School of Science, FI-00076 Aalto, Finland}
\author{G. Spada\orcidlink{0000-0002-3505-8603}}
\affiliation{
	School of Science and Technology, Physics Division, Universit\`a di Camerino, 62032 Camerino, Italy}
\affiliation{
	INFN, Sezione di Perugia, I-06123 Perugia, Italy}
\author{S. Giorgini\orcidlink{0000-0001-9146-7025}}
\affiliation{Pitaevskii BEC Center, CNR-INO and Dipartimento di Fisica, Universit\`a di Trento, I-38123 Trento, Italy}
\author{S. Pilati\orcidlink{0000-0002-4845-6299}}
\affiliation{
	School of Science and Technology, Physics Division, Universit\`a di Camerino, 62032 Camerino, Italy}
\affiliation{
	INFN, Sezione di Perugia, I-06123 Perugia, Italy}
\author{A. Recati\orcidlink{0000-0002-8682-2034}}
\affiliation{Pitaevskii BEC Center, CNR-INO and Dipartimento di Fisica, Universit\`a di Trento, I-38123 Trento, Italy}

\title{Supersolid phase in two-dimensional soft-core bosons at finite temperature}

	\begin{abstract}
    
        The supersolid phase of soft-core bosons in two dimensions is investigated using the self-consistent Hartree-Fock and quantum Monte Carlo methods. An approximate phase diagram at finite temperatures is initially constructed using the mean-field approach, which is subsequently validated through precise path-integral simulations, enabling a microscopic characterization of the various phases. Superfluid and melting/freezing transitions are analyzed through the superfluid density and the long-range behavior of correlation functions associated with positional and orientational order, in accordance with the general picture of Berezinskii-Kosterlitz-Thouless transitions. A broad region at low temperatures is identified where the supersolid phase exists, separating the uniform superfluid phase from the normal quasi-crystal phase. Additionally, a potential intermediate hexatic phase with quasi long-range orientational order is identified in a narrow region between the normal solid and fluid phases. These findings establish self-consistent Hartree-Fock theory beyond the local density approximation as an effective tool, complementary to computationally intensive quantum Monte Carlo simulations, for investigating the melting of the supersolid phase and the possible emergence of the hexatic superfluid phase in bosonic systems with various interaction potentials.
        \end{abstract}

	\maketitle
	
\begin{figure*}
\includegraphics{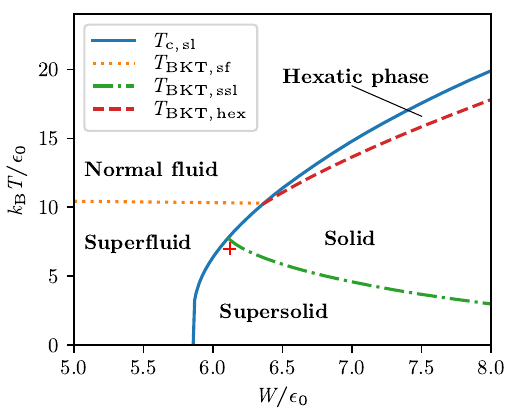}
\includegraphics{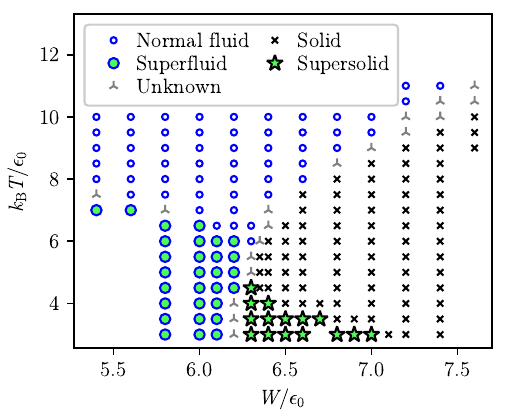}
\caption{\label{fig:phase_diagram}  
Phase diagram of 2D soft-core bosons as a function of temperature $T$ and strength $W$ of the interaction potential~\eqref{eq:soft_core}. The density is fixed at $\rho R^2 = 4.4$, with $R$ the range of the soft-core interaction potential~\eqref{eq:soft_core}. The parameter $\epsilon_0 = \hbar^2/(2mR^2)$ is taken as the energy scale. Left panel: phase diagram obtained within the self-consistent Hartree-Fock approximation. The density profiles of the condensate and thermal fractions shown in Fig.~\ref{fig:density_profile} are taken at the point marked by the red cross. Other numerical parameters are: discretization parameter $M = 22$, supercell linear size $L = 6$, see Appendices~\ref{app:momentum_space}-\ref{app:iterative_solution}  for details on the numerical implementation of the HF approximation and Section~\ref{sec:HF_analysis} for the explanation of how the phase boundaries have been obtained. Right panel: phase diagram obtained from PIMC simulations with $N=4032$ particles, see Section~\ref{sec:PIMC} for details. Points are classified by analyzing the correlation functions $G_6(r)$ and the superfluid response. \qql Normal\qqr means here non-superfluid (zero superfluid density in the thermodynamic limit).} 
\end{figure*}

\section{Introduction}

Supersolidity is a peculiar state of matter where both  $\mathrm{U(1)}$ and translational continuous symmetries in $d$ dimensions are spontaneously broken. According  to  hydrodynamic theory \cite{Andreev1969} for a supersolid in $d$ dimensions $1+d$ distinct gapless dispersive modes, respectively Bogoliubov and crystal phonons, characterise the spectrum of the system. These modes correspond to the Nambu-Goldstone bosons emerging due to the broken symmetries, as shown by generalising the
Goldstone theorem  to non-Lorentz-invariant systems \cite{Watanabe2012}. 

In three dimensions (3D), supersolids feature long-range order (LRO) in the density fluctuations, which share the periodicity of the lattice structure, and off-diagonal long-range order (ODLRO) in the one-particle density matrix as a result of phase coherence and Bose-Einstein condensation (BEC). In two dimensions (2D), both types of order parameter are not stable against thermal fluctuations, but a similar classification of the supersolid phase can be defined in terms of power-law as opposed to exponential decay of the corresponding correlation functions~\cite{Boninsegni2012}.  The transitions associated to the emergence of quasi-long range order, namely correlation functions with algebraic decay, are generally of the Berezinskii-Kosterlitz-Thouless (BKT) type.

From the experimental point of view, first inconclusive evidences of supersolid behavior were observed in crystals of $^4$He~\cite{Kim2004a,Kim2004}. The interpretation of these results was not straightforward~\cite{Balibar2010} and more careful studies have finally shown that solid helium does not exhibit a finite superfluid response~\cite{Kim2012}. This conclusion agrees with the general belief that superfluidity in crystals is ruled out if the number of atoms is commensurate with the number of lattice sites~\cite{Prokofev2005}.

Unambiguous evidence of supersolidity was instead reported in ultracold dipolar gases, where thousands of atoms form phase coherent clusters. The cluster-type supersolid was proposed long ago by E. P. Gross \cite{Gross1960}. 
First experiments realised linear arrays of clusters ~\cite{Tanzi2019,Bottcher2019,Chomaz2019}, and observed the distinct Goldstone modes associated with the two spontaneously broken symmetries~\cite{Tanzi2019b,Guo2019,Natale2019}. More recently, planar configurations have been realized~\cite{Norcia2021,Bland2022} and quantized vortices have been created \cite{Casotti2024}. New experiments are focusing on the role of temperature in suppressing coherence between droplets~\cite{Ferlaino2021,Baena2023} opening the way to the study of the finite temperature phase diagram.
On the other hand, a supersolid phase in the sense predicted by Andreev and Lifshitz~\cite{Andreev1969}, namely a condensate of zero-point vacancies, has been observed in  a single layer of $^3\mathrm{He}$ adsorbed on a carbon nanotube~\cite{Todoshchenko2022,Todoshchenko2024}.

On the theory side, while usually ultracold dilute Bose gases are very well described by the Gross-Pitaevskii (GP) theory, the dipolar cluster and supersolid states require the inclusion of the repulsive Lee-Huang-Yang correction in the GP equation (a.k.a. extended GP equation) to stabilize the attractive component of the anisotropic dipolar interaction~\cite{Lima2012,Wachtler2016}. 
Quantum droplets of dipolar atoms have also been investigated beyond the extended GP scheme using quantum Monte-Carlo methods~\cite{Saito2016,Macia2016,Bottcher2019a}.

Concerning the study of phase transitions the dipolar gas platform has a drawback: in order to avoid the collapse of the gas due to the head-to-tail dipole-dipole attraction, a confinement in the direction of the atomic dipole moment is necessary. The system is eventually composed of mesoscopic elongated clusters organised in a 2D structure. These features make a scaling to the thermodynamic limit and the study of reduced dimensionality not straightforward. 

From a theoretical point of view, a very useful model, which exhibits a supersolid phase in 2D and in 3D is a Bose gas with a soft-core interaction potential (see, e.g.,~\cite{Pomeau1994,Boninsegni2012b})
\begin{equation}
		\label{eq:soft_core}
		V(\vb{r}) = W\theta(R -\abs{\vb{r}})\,,	
	\end{equation}
with $\vb{r}$ the vector connecting two atoms, $\theta(x)$ the Heaviside step function, $R$ the range of the potential, which will serve as our unit of length, and $W > 0$ a constant determining the strength of the interaction.
For later convenience, we also define the unit of energy for the present study as $\epsilon_0 = \hbar^2/(2mR^2)$. 	
Interactions in the soft-core model are purely repulsive, resulting in  a well-defined equilibrium state in the thermodynamic limit.
 
In 2D this model has been studied at zero temperature by using GP theory \cite{Pomeau1994} and path-integral Monte Carlo (PIMC), focusing on the low temperature limit \cite{Saccani2011,Saccani2011a}. The model features a normal cluster solid phase, a cluster supersolid phase, and a uniform superfluid phase. The zero-temperature excitation spectrum of the various phases has also been explored~\cite{Macri2013,Saccani2012,Poli2024}.

As already mentioned, phase transitions in 2D have peculiar features due to the enhanced role of thermal fluctuations. The normal to superfluid transition is of the BKT type and is signaled by the universal jump of the superfluid density~\cite{Nelson1977}. Different scenarios can instead apply to the freezing/melting transition in 2D. First of all, a 2D crystal can sustain only positional quasi-LRO, meaning that the pair correlation function along a symmetry axis exhibits a power-law decay. Orientational LRO is instead present resulting in a finite order parameter. In classical systems the melting to the fluid state, where both positional and orientational order decay exponentially, may occur as a two-step process involving the hexatic phase as an intermediate state~\cite{Kapfer2015}. In the hexatic phase positional correlations decay exponentially, while orientational correlations decay as a power law. Depending on the type of interaction potential, one could have that the liquid/hexatic transition is first order while the hexatic/crystal transition is of the BKT type~\cite{Kapfer2015}. Alternatively, both transitions are of the BKT type according to the celebrated Kosterlitz-Thouless-Halperin-Nelson-Young (KTHNY) theory~\cite{Halperin1978,Nelson1979,Young1979}. Evidences of the hexatic phase and of the KTHNY melting scenario have been observed in systems of colloidal particles~\cite{Zahn1999,Gasser2010}.   
A different scenario predicts a first-order liquid to solid transition without hexatic phase~\cite{Strandburg1988,Simionesco2006}.

More complications arise when one adds quantum-mechanical effects to the melting scenario. Intriguing question are: a) Are quantum exchange effects relevant for the 2D freezing/melting transition? b) Is there an hexatic phase and could this phase sustain a finite superfluid response? c) Is the superfluid to supersolid transition of the first order or of the BKT type? Concerning questions a) and b), a theoretical study of dipoles in 2D including quantum effects argued that the hexatic phase could survive down to very low temperatures well within the degenerate regime~\cite{Bruun2014}. As for question c), at $T=0$ one expects that the energy of the crystal state of the soft-core model becomes lower than the homogeneous superfluid for densities above the critical point. However, at finite $T$, it is unclear whether the transition remains first order.

In this article we investigate theoretically the finite temperature phase diagram of soft-core bosons in 2D and we establish quantitatively the boundaries of the various phases in the thermodynamic limit. We use both a Hartree-Fock (HF) scheme, where solutions of the GP equation and thermal excitations are treated self-consistently without resorting to local density approximation, and exact PIMC simulations to validate the main results of the HF analysis and to characterize the different phases through the calculation of the relevant correlation functions. 

We find that the supersolid phase occupies a significant region of the phase diagram at large interaction strength and low temperatures. The hexatic phase might exist in a thin region between the crystal and the fluid, but there appears to be little room for a hexatic superfluid phase~\cite{Mullen1994}. Furthermore, PIMC simulations indicate that the transition from the superfluid to the supersolid is accompanied by a jump in the superfluid density and in the orientational order parameter.
In the regime of large interaction strength, we also find that the orientational order parameter increases with temperature, reaches a maximum and then rapidly vanishes as the system enters the normal fluid phase. We attribute this counterintuitive behavior to an enhanced role of quantum fluctuations.

The structure of the paper is as follows. The main results of the present paper, the phase diagrams obtained, respectively, from HF theory and PIMC simulations, are presented and compared in Sec.~\ref{sec:phase-diagram}.  In Sec.~\ref{sec:HF_variational}, we present in detail the self-consistent HF scheme used to obtain the phase diagram. The critical temperature of the first order fluid-solid transition predicted by HF theory is obtained by computing the free energies of the uniform and density-modulated phases, as explained in Sec.~\ref{sec:phase_diagram_HF}. Using HF theory, we calculate the superfluid density with the phase twist method and also the crystal's bulk and shear modulus. In Sec.~\ref{sec:BKT_transitions}, these are used together with the universal relations of the BKT transition to estimate the  critical temperatures of the transitions from superfluid to normal and from the crystal to the hexatic phase. In Sec.~\ref{sec:PIMC} we briefly describe the PIMC method and we discuss results for the superfluid density, the orientational order parameter and the long-range behavior of the pair correlation function and of the orientational correlation function by varying temperature and strength of the interaction. Final conclusions are examined at the end in Sec.~\ref{sec:conclusion}. Technical details and additional results for the HF method are collected in Appendices~\ref{app:self-consistent}-\ref{app:bulk_shear}. Finite size effects in PIMC simulations are also further discussed in Appendix~\ref{app:finite_size_pimc}.

\section{Phase diagram of soft-core bosons}
\label{sec:phase-diagram}

In this section we discuss the phase diagram of 2D soft-core bosons at finite temperature as obtained from HF self-consistent calculations and exact PIMC simulations. The two phase diagram are shown in Fig.~\ref{fig:phase_diagram}, where the following phases are identified:  First we distinguish between a normal fluid and a superfluid, both occurring in the regime of relatively weak interaction strength, at high and low temperature respectively. In 2D the superfluid transition is of the BKT type and entails a universal jump of the superfluid density from a finite critical value to zero. Both within the HF framework and in PIMC simulations, we calculate the superfluid density $\rho_S$ and use the Nelson-Kosterlitz criterium (see Eq.~(\ref{eq:KN_relation}) below) to determine whether the system is in the superfluid or in the normal phase. 

Notice that the HF theory can not describe the BKT superfluid transition and would predict a finite $\rho_S$ even above the BKT transition temperature. On the contrary, PIMC simulations are known to reliably reproduce the critical behavior of $\rho_S$ at the transition point~\cite{Pilati2008} once finite-size scaling is properly accounted for. Nevertheless, the above simple criterium provides an accurate estimate of the transition temperature if the size of the simulation is large enough. In the comparison between HF and exact phase diagram, we notice that HF theory overestimates the extension of the superfluid region up to higher temperatures ($k_BT/\epsilon_0\sim10$, while PIMC simulations indicate the transition at $k_BT/\epsilon_0\sim7$ for the smallest interaction strength). Furthermore, whereas the superfluid to normal transition point appears to be almost insensitive to interactions within the HF scheme, exact simulations show a decrease of the transition temperature with increasing interaction strength $W$.

By increasing the strength $W$ of the interaction at relatively low temperature one enters the supersolid phase, characterized by a finite superfluid density (above the universal critical value set by BKT theory) and a density modulation corresponding to a triangular lattice. Within the HF approach, the spatially modulated solution corresponds to a lower free energy compared to the uniform solution and the lattice constant is also be determined via a minimization procedure. PIMC simulations, instead, provide direct access to the global orientational order parameter $|\psi_6|^2$ defined in Sec.~\ref{sec:PIMC} from the integral of the orientational correlation function $G_6(r)$. The order parameter $|\psi_6|^2$ remains finite in the quasi-crystal phase, even though the density-density correlation function exhibits peaks which decay with a power law according to the theory of the crystal phase in 2D. Simulations seem to indicate an abrupt appearance of the orientational order parameter as a function of the interaction strength $W$ (see Fig.~\ref{fig:pimc_order_param}). A discontinuous jump of $|\psi_6|^2$ would be compatible both with a first-order and a BKT-type transition. However, much larger simulations would be needed to exclude a continuous transition or the existence of an intermediate hexatic phase between the fluid and the solid. We also notice that the HF prediction of the onset of supersolidity as a function of interaction is in reasonable quantitative agreement with the results of PIMC simulations.

The transition from supersolid to solid is determined using the criterium of the superfluid density $\rho_S$ analogously to the normal to superfluid transition in the uniform phase. Also in this case HF theory overestimates the extension of the supersolid region, while the decrease of the transition temperature with interaction strength is in agreement with the PIMC phase diagram. It is commonly accepted that the transition between the supersolid and the solid phase persists down to zero temperature (see, e.g. \cite{Saccani2012,Macri2013}), however an exact zero-temperature analysis is still lacking. Interestingly, the presence of such zero-temperature critical point has been argued to lead to a renormalisation of the Kosterlitz-Nelson criterium, as predicted for the quantum phase model \cite{Doniach1981}. 
A simple linear extrapolation to zero temperature of the normal to supersolid transition point based on the phase diagram in Fig.~\ref{fig:phase_diagram}, yields an estimate $W/\epsilon_0\approx 9$ of the quantum critical point in agreement with Ref.~\cite{Macri2013}.

The HF phase diagram includes an hexatic phase separating the normal fluid from the normal solid. This is signaled by a spatially modulated solution of the HF equations being lower in free energy compared to the uniform solution and a Young elasticity modulus $Y$ being smaller than the critical value corresponding to the universal jump of the BKT theory of 2D melting. Indeed, the KTHNY theory predicts at this value of $Y$ a transition from a quasi-crystal to an hexatic phase where translational order is short ranged and orientational order decays algebraically yielding a vanishing global order parameter $|\psi_6|^2$. According to the HF phase diagram, the hexatic phase can only be accessed from the normal fluid. This suggests the absence of an hexatic superfluid phase within HF theory. Using PIMC simulations we tried to look carefully for the hexatic phase especially in the region of high temperatures and large interactions strengths. Unfortunately, this phase is not associated with a global order parameter and can be recognized only by studying the behavior of the density-density and orientational correlation functions which should decay, respectively, exponentially and algebraically. This is difficult to establish in a firm way for the system sizes allowed by our simulations. However, some indication of hexatic behavior are visible in the upper right corner of the phase diagram in Fig.~\ref{fig:phase_diagram} corresponding to the points labeled as "unknown" which we can not clearly assign neither to the homogeneous normal phase nor to the solid phase. Similar "unknown" points are also present in the phase diagram of Fig.~\ref{fig:phase_diagram} in a narrow slice separating the superfluid and the supersolid or normal solid phase. For these points the orientational order parameter $|\psi_6|^2$ shows significative size dependence and the corresponding correlation function $G_6(r)$ decays slowly. We can not conclude whether these points lie very close to the transition and suffer of large finite-size effects or they are associated with the existence of an intermediate hexatic phase which, in any case, would occupy a very thin region of the phase diagram. 

In Sec.~\ref{sec:HF_analysis} and Sec.~\ref{sec:PIMC}, more details on the analysis leading to the results shown in Fig.~\ref{fig:phase_diagram} are provided.

\section{Self-consistent Hartree-Fock theory analysis}
\label{sec:HF_analysis}

The first method employed to determine the phase diagram shown in Fig.~\ref{fig:phase_diagram} is self-consistent HF theory. HF theory represents the simplest form of independent-particle approximation for quantum many-body systems and serves as the foundation for more advanced methods in nuclear and atomic structure theory, as well as in quantum chemistry. It is widely utilized for fermionic systems, where it underpins the intuitive understanding of atomic and molecular energy levels in terms of single-particle spin-orbital occupations.

For bosonic systems, HF theory has been applied in its fully self-consistent form to investigate the thermodynamics of ultracold atoms confined in harmonic traps~\cite{Giorgini1997,Holzmann1999}. However, in these studies, the solution of the HF equations relied on the use of the semiclassical or local density approximation which greatly simplifies the numerics. The local density approximation combined with Bogoliubov theory has also been employed to map out the finite temperature phase diagram of an ultracold dipolar quantum gas, leading to the remarkable finding that the supersolid phase appears out of a uniform superfluid upon increasing temperature~\cite{Baena2023}.

A key contribution of the present work is demonstrating the practical feasibility of obtaining self-consistent HF solutions for bosonic systems at finite temperature beyond the local density approximation scheme. As detailed in Sec.~\ref{sec:HF_variational}, HF theory provides a natural extension of the GP equation to finite temperatures, offering a conceptually clear and computationally efficient framework. Furthermore, it is shown in Fig.~\ref{fig:phase_diagram} that the phase diagram predicted by HF theory is in qualitative agreement with results from computationally intensive PIMC simulations. 

In Sec.~\ref{sec:HF_variational}, the self-consistency equations of HF theory for bosons are derived using a finite temperature variational principle. The variational nature of the HF approximation even at finite temperature is rarely emphasized, however it is important here since the solid phase is identified by checking when its free energy becomes lower than the one of the homogeneous phase, as detailed in Sec.~\ref{sec:phase_diagram_HF}. Then, in Sec.~\ref{sec:BKT_transitions} it is explained how the calculations of the superfluid density and of the Young modulus, in connection with the use of the universal Nelson-Kosterlitz relations, allow one to obtain rather accurate estimates of the BKT critical temperatures, respectively, of the normal-superfluid transition and of the solid-hexatic transition. Details on the derivation and the numerical solution of the self-consistency equations of HF theory are provided in Appendices~\ref{app:self-consistent}-\ref{app:bulk_shear}.

	\subsection{Hartree-Fock theory  from the finite temperature variational principle}
	\label{sec:HF_variational}
    
	Consider the Hamiltonian $\ham = \ham_{\rm free} + \ham_{\rm int}  -\mu \hat{N}$ of an interacting two-dimensional Bose gas, which consists of the sum of a free or non-interacting term
	\begin{equation}
		\label{eq:H_free_first}
		\ham_{\rm free}= \int \dd{\vb{r}} \hat{\psi}^\dg(\vb{r})\qty(-\frac{\hbar^2\nabla^2_{\vb{r}}}{2m} + U(\vb{r}))\hat{\psi}(\vb{r})\,,
	\end{equation}
	an interaction term of the density-density type
	\begin{equation}
		\label{eq:H_int_first}
		\ham_{\rm int} =  \frac{1}{2}\iint\dd{\vb{r}}\dd{\vb{r}'}V(\vb{r}-\vb{r}')\hat{\psi}^\dg(\vb{r})\hat{\psi}^\dg(\vb{r}') \hat{\psi}(\vb{r}')\hat{\psi}(\vb{r})\,,
	\end{equation}
	and a chemical potential $\mu$ term proportional to the total particle number operator 
	\begin{equation}
		\label{eq:N_op_def}	
		\hat{N} = \int \dd{\vb{r}} \hat{\psi}^\dg(\vb{r})\hat{\psi}(\vb{r})\,,
	\end{equation}
	In~\eqref{eq:H_free_first}-\eqref{eq:N_op_def} $\hat\psi(\vb{r})$ is a bosonic field operator satisfying the standard commutation relations
	\begin{gather}
		\label{eq:comm_rel}
		[\hat{\psi}(\vb{r}),\hat{\psi}^\dg(\vb{r}')] = \delta(\vb{r}-\vb{r}')\,,\\ 
		[\hat{\psi}(\vb{r}),\hat{\psi}(\vb{r}')] = [\hat{\psi}^\dg(\vb{r}),\hat{\psi}^\dg(\vb{r}')] = 0\,,
	\end{gather}
	$U(\vb{r})$ a single-particle potential and $V(\vb{r})$  the interaction potential. In particular, we are interested in the soft-core interaction potential~\eqref{eq:soft_core}. It was shown in Ref.~\cite{Pomeau1994} using the GP equation, that this type of interaction leads to a supersolid phase at zero temperature.
	
	A particularly neat way to obtain the equations of HF theory is using the finite temperature variational principle for the thermodynamic grand potential, based on the Bogoliubov inequality~\cite{Feynman1998,Blaizot1986,Peotta2022,Tam2024}
	\begin{equation}
		\label{eq:Bogoliubov_inequality}
		\Omega \leq \Omega_0 + \ev*{\ham -\ham_0} = \Omega_{\rm m.f.}\,,	
	\end{equation}
	where $\ham$ is the many-body Hamiltonian just introduced, while $\ham_0$ is an auxiliary Hamiltonian. The corresponding grand potentials $\Omega$ and $\Omega_0$ are defined by
	\begin{equation}
		\label{eq:Omega_def}
		\Omega_{(0)} = -\frac{1}{\beta}\ln \Tr \qty\big[e^{-\beta \ham_{(0)}}]\,.
	\end{equation}
	The expectation value on the right-hand side of~\eqref{eq:Bogoliubov_inequality} is evaluated with respect to the statistical ensemble given by $\ham_0$, namely
	\begin{equation}
		\label{eq:H-H0_ev}
		\ev*{\ham -\ham_0} = \Tr\qty\big[\hat{\rho}_0\qty\big(\ham -\ham_0)]\,.	
	\end{equation}
	with $\hat{\rho}_0 = e^{-\beta \ham_0}/\Tr\qty\big[e^{-\beta \ham_0}]$.
	In the following, we adopt the convention that all expectation values are evaluated with respect to the variational Hamiltonian $\ham_0$ as above.
	
	The Hamiltonian $\ham_0$ is called variational because it is chosen so as to minimize the right-hand side of~\eqref{eq:Bogoliubov_inequality}, called the mean-field grand potential $\Omega_{\rm m.f.}$, thus giving the best possible approximation to the exact grand potential $\Omega$. This is equivalent to minimizing with respect to the density matrix $\hat\rho_0$. 
	
	The HF approximation is obtained by taking $\ham_0$ to be an operator at most quadratic in $\hat{\psi}(\vb{r}),\,\hat{\psi}^\dg(\vb{r})$.
	In this work, $\ham_0$ takes the form
	\begin{equation}
		\label{eq:H0}
		\begin{split}
			\ham_0 &= \int \dd{\vb{r}} \hat{\psi}^\dg(\vb{r})\qty(-\frac{\hbar^2 \grad_{\vb{r}}}{2m}+ U(\vb{r})-\mu)\hat{\psi}(\vb{r}) 
			\\
			&\quad+ \int \dd{\vb{r}} \qty\big[\xi(\vb{r})\hat{\psi}^\dg(\vb{r})+\xi^*(\vb{r})\hat{\psi}(\vb{r})] \\
			&\quad+\iint \dd{\vb{r}} \dd{\vb{r}'} \Gamma(\vb{r},\vb{r}')\hat{\psi}^\dg(\vb{r})\hat{\psi}(\vb{r}')\,.
		\end{split}
	\end{equation}
	The fields $\xi(\vb{r})$ and $\Gamma(\vb{r},\vb{r}')$ should be understood as variational parameters. The HF potential $\Gamma(\vb{r},\vb{r}')$ satisfies the constraint $\Gamma(\vb{r},\vb{r}') = \Gamma^*(\vb{r}',\vb{r})$ imposed by the Hermiticity of $\ham_0$. It is also possible to include a pairing term of the form $\Delta(\vb{r},\vb{r}')\hat{\psi}^\dg(\vb{r}) \hat{\psi}^\dg(\vb{r}') + \text{h.c.}$, leading to the Hartree-Fock-Bogoliubov (HFB) approximation. However, we have found that it is significantly more difficult to obtain a self-consistent solution within HFB theory, at least with the iterative algorithm presented in Appendix~\ref{app:iterative_solution}. Therefore only HF theory is discussed here. 
	
	Due to the presence of the linear term in the second line of~\eqref{eq:H0}, the variational Hamiltonian is not number conserving and the field operator can acquire a nonzero expectation value, which is called the condensate wave function and denoted by $\psi(\vb{r}) = \ev*{\hat{\psi}(\vb{r})}$. Notice that in 2D the squared order parameter $|\psi(\vb{r})|^2$ should be interpreted as the quasi-condensate density (see e.g. Ref.~\cite{Capogrosso-Sansone2010}), which provides a legitimate description of the thermodynamic state at temperatures much smaller than the degeneracy temperature $T^\ast=2\pi\hbar^2\rho/k_Bm$ set by the density $\rho$.
	It is useful to introduce the fluctuation operator
	\begin{equation}
		\label{eq:fluc_op}
		\widetilde{\psi}(\vb{r}) \overset{\rm def}{=} \hat{\psi}(\vb{r}) - \ev*{\hat{\psi}(\vb{r})} = \hat{\psi}(\vb{r}) - \psi(\vb{r})\,.
	\end{equation}
	The following properties are an immediate consequence of the definition of fluctuation operator
	\begin{gather}
		\label{eq:fluc_prop_1}
		\ev*{\widetilde{\psi}(\vb{r})} = \ev*{\widetilde{\psi}^\dg(\vb{r})}  = 0\,,\\ 
		\label{eq:fluc_prop_2}
		\ev*{\hat{\psi}^\dg(\vb{r})\hat{\psi}(\vb{r}')} = \psi^*(\vb{r})\psi(\vb{r}')+ \ev*{\widetilde{\psi}^\dg(\vb{r})\widetilde{\psi}(\vb{r}')} \,.
	\end{gather}

	By imposing that the functional derivatives of the mean-field grand potential $\Omega_{\rm m.f.}$
	with respect to the variational parameters vanish (for details see Appendix~\ref{app:self-consistent}), one obtains the extended GP equation
	\begin{gather}
		\label{eq:extGP_cont_1}
		\begin{split}
			&\mu \psi(\vb{r}) = 	\qty(-\frac{\hbar^2 \grad_{\vb{r}}^2}{2m}+U(\vb{r}))\psi(\vb{r}) \\
			&+ \int \dd{\vb{r}'} V(\vb{r}-\vb{r}')\qty\Big[\abs{\psi(\vb{r}')}^2+\ev*{\widetilde{\psi}^\dg(\vb{r}')\widetilde{\psi}(\vb{r}')}]\psi(\vb{r})\\
			&+\int \dd{\vb{r}'} V(\vb{r}-\vb{r}')\ev*{\fpsi^\dg(\vb{r}')\fpsi(\vb{r})}\psi(\vb{r}')\,,
		\end{split}
	\end{gather}
	and the self-consistency equation for $\Gamma(\vb{r},\vb{r}')$
	\begin{gather}
		\label{eq:sc_cont_1}
		\Gamma(\vb{r},\vb{r}') = \Gamma_{\rm H}(\vb{r},\vb{r}') + \Gamma_{\rm F}(\vb{r},\vb{r}')\,,
		\\
		\label{eq:sc_cont_2}
		\begin{split}
			&\Gamma_{\rm H}(\vb{r},\vb{r}') = \delta(\vb{r}-\vb{r}')\\
			&\hspace{0.5cm}\times\int \dd{\vb{r}''}V(\vb{r}-\vb{r}'') \ev*{\hat{\psi}^\dg(\vb{r}'')\hat{\psi}(\vb{r}'')}\,,
		\end{split}
		\\
		\label{eq:sc_cont_3}
		\Gamma_{\rm F}(\vb{r},\vb{r}') = V(\vb{r}-\vb{r}')\ev*{\hat{\psi}^\dg(\vb{r}')\hat{\psi}(\vb{r})}\,. 
	\end{gather}
	Note that we have split the Hartree-Fock potential into its Hartree  $	\Gamma_{\rm H}(\vb{r},\vb{r}')$ and Fock $\Gamma_{\rm F}(\vb{r},\vb{r}')$ components. The HF approximation consists in finding a solution of the system of nonlinear equations~\eqref{eq:extGP_cont_1}-\eqref{eq:sc_cont_3}. Note that the HF potential enters in the extended GP equation through the expectation values of fluctuation operators since these are evaluated on the statistical ensemble given by the variational Hamiltonian $\ham_0$. At low enough temperature the expectation values of the fluctuation operators in~\eqref{eq:extGP_cont_1} can be neglected and one obtains the standard GP equation in its static form, which is an equation for the condensate wave function alone.

	\begin{figure}
		\includegraphics{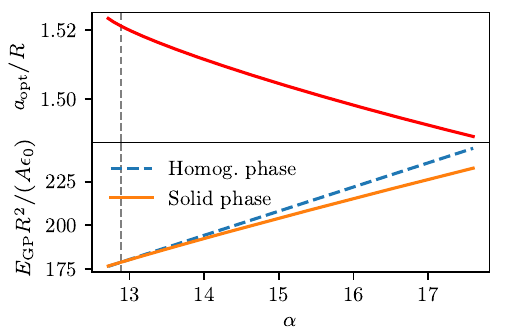}
		\caption{\label{fig:lattice_constant}Upper panel: optimal lattice constant $a_{\rm opt}$ as a function of the dimensionless interaction parameter $\alpha$~\eqref{eq:alpha} at $T=0$. Lower panel: GP energy functional per unit area $E_{\rm GP}/A$~\eqref{eq:GP_functional_zero_T} for the homogeneous phase and the solid (crystalline) phase. For each $\alpha$,  the energy functional is computed using the optimal lattice constant given in the upper panel. The first order  transition from the homogeneous to the solid phase occurs at the energy crossing denoted by the vertical dashed line ($\alpha_{\rm c} = 12.9$ and $a=1.521$ with discretization parameter $M =22$, see Eq.~\eqref{eq:a_Mb} in Appendix~\ref{app:discretization}).}
	\end{figure}
	
	\subsection{Homogeneous vs spatially modulated solution}
	\label{sec:phase_diagram_HF}

	In this subsection we examine, within HF theory, the transition from a homogeneous phase to the state with spatial density modulations indicating spontaneously broken translational symmetry.
	Details regarding the numerical solution of the self-consistency equations~\eqref{eq:extGP_cont_1}-\eqref{eq:sc_cont_3} are collected in Appendices~\ref{app:momentum_space}-\ref{app:iterative_solution}.
	
	The many-body Hamiltonian $\ham$ is translational invariant if the external potential is constant, $U(\vb{r})=\text{const}$. Continuous translational symmetry can be spontaneously broken by spatial modulations of the density. The residual discrete translational symmetry is characterized by a  pair of fundamental lattice vectors $\vb{a}_{i=1,2}$ entering in the periodicity constraints
	\begin{gather}
		\label{eq:psi_cond_per}
		\psi(\vb{r}) = \psi(\vb{r}+ \vb{a}_i)\,, \\
		\label{eq:Gamma_cond_per}
		\Gamma(\vb{r}, \vb{r}') = \Gamma(\vb{r}+ \vb{a}_i, \vb{r}'+ \vb{a}_i)
		 \qq{for} i = 1,2\,.
	\end{gather}	
	Since the lattice structure is formed spontaneously, the vectors $\vb{a}_i$ are themselves variational parameters.
	
	\begin{figure}
		\includegraphics{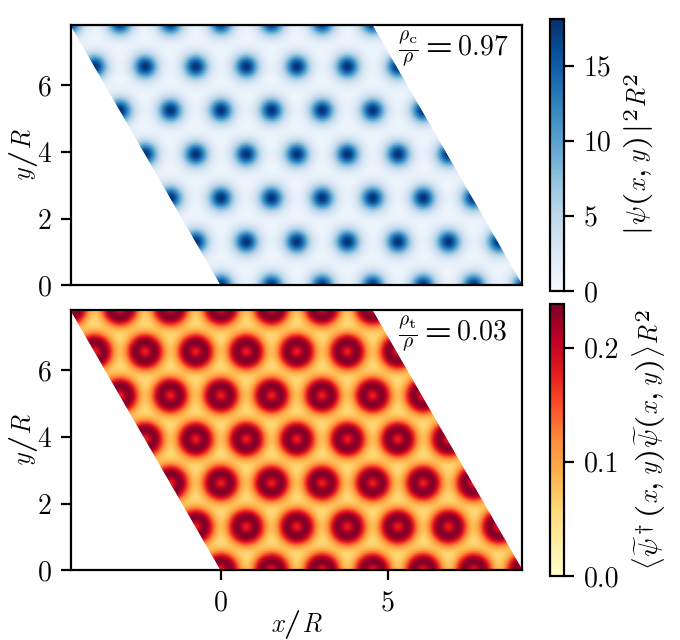}
		\caption{\label{fig:density_profile} Density profiles of condensate $\abs{\psi(\vb{r})}^2$ (top) and thermal excitations $\ev*{\fpsi^\dg(\vb{r})\fpsi(\vb{r})}$ (bottom) obtained from HF calculations. Parameters are $k_{\rm B}T = 7\epsilon_0$, $W = 6.12\epsilon_0$, $\rho R^2=4.4$, $a_{\rm opt} = 1.516$, $(M, L)=(22, 6)$, where $M$ is the discretization parameter and $L$ the number of unit cells along each of the vectors~$\vb{a}_i$, see Appendices~\ref{app:momentum_space}-\ref{app:iterative_solution}. $\rho_{\rm c}$ and $\rho_{\rm t}$ are the average densities of condensate and thermal excitations, respectively and $\rho = \rho_{\rm c}+\rho_{\rm t}$ the total density. The point in the phase diagram at which the density profiles are calculated is close to the first order transition temperature $T_{\rm c,sl}$ and is marked as a cross in Fig.~\ref{fig:phase_diagram}.}
	\end{figure}
	
	At $T=0$ the expectation values $\ev*{\fpsi^\dg(\vb{r})\fpsi(\vb{r}')}$ vanish and one is left with the task of finding a solution of the GP equation alone.  In the case of the soft-core potential~\eqref{eq:soft_core}, the solution of the GP equation depends on the single dimensionless quantity~\cite{Macri2013}
	\begin{equation}
		\label{eq:alpha}
		\alpha = \frac{mW\rho R^4}{\hbar^2} = \frac{W\rho R^2}{2\epsilon_0}\,.	
	\end{equation}
	At zero temperature, the total density $\rho$ is equal to the condensate density $\rho_{\rm c}=\int \dd{\vb{r}}\abs{\psi(\vb{r})}^2/A$, with $A$ the area. 	
	
	Above a critical value $\alpha_{\rm c}$, a solution with the symmetry of a triangular lattice is favoured with respect to the homogeneous phase~\cite{Pomeau1994,Macri2013}. The precise critical value depends on the lattice constant $a = \abs{\vb{a}_i}$ of the triangular lattice. The optimal value $a_{\rm opt}$ of the lattice constant is calculated by 
 	minimizing 	the GP energy functional per unit area $E_{\rm GP}/A$ as a function of  $a$ at fixed density $\rho$ whenever a modulated solution exists. The GP energy functional $E_{\rm GP}$ is defined by
 	\begin{equation}
 		\label{eq:GP_functional_zero_T}
 		\begin{split}
 			&E_{\rm GP}[\psi(\vb{r}), \psi^*(\vb{r})] \\ &= \int \dd{\vb{r}} \psi^*(\vb{r})\qty(-\frac{\hbar^2 \nabla^2_{\vb{r}}}{2m}+U(\vb{r}))\psi(\vb{r})  \\
 			&+ \int\int \dd{\vb{r}}\dd{\vb{r}'}V(\vb{r}-\vb{r}')\frac{1}{2}\abs{\psi(\vb{r})}^2\abs{\psi(\vb{r}')}^2\,.
 		\end{split}
 	\end{equation}
 	The optimal lattice constant at zero temperature as a function of $\alpha$ is shown in Fig.~\ref{fig:lattice_constant}. Using the discretization procedure presented in Appendix~\ref{app:discretization}, we find the critical value $\alpha_{\rm c}=12.9$, slightly above $\alpha_{\rm c} = 12.7$ obtained in Ref.~\cite{Macri2013} with a Gaussian ansatz and closer to the path integral quantum Monte Carlo $\alpha_{\rm c} = 13.4$ prediction at density $\rho R^2 = 4.4$ from the same reference.

\begin{figure}
	\includegraphics{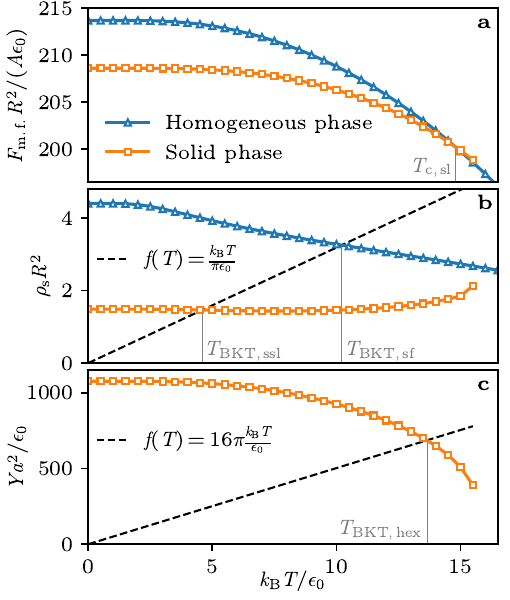}
	\caption{\label{fig:three_panels} Determination of the transition temperatures $T_{\rm c, sl}$ and $T_{\rm BKT,(sf,ssl,hex)}$ shown in the phase diagram of Fig.~\ref{fig:phase_diagram}: (\textbf{a}) the first order transition between the solid (crystalline) phase and the homogeneous fluid phase occurs at the temperature $T_{\rm c,sl}$ at which the corresponding mean-field free energies~\eqref{eq:free_energy_def} are equal; (\textbf{b}) superfluid density $\rho_{\rm s}$  in the homogeneous and in the solid phases obtained from a quadratic fit of the mean-field free energy according to~\eqref{eq:sf_rho_def}. The straight dashed line corresponds to the right-hand side of~\eqref{eq:KN_relation} in appropriate units, thus the crossings with the superfluid density curves give an estimate of the critical temperatures of the superfluid BKT transition within the homogeneous ($T_{\rm BKT,sf}$) and solid ($T_{\rm BKT,ssl}$) phases; (\textbf{c})  Young modulus as a function of temperature. In the same way as in panel \textbf{b}, the intercept with the straight line determines the critical temperature of the BKT transition from the solid to the hexatic phase according to the universal relation for two-dimensional melting~\eqref{eq:KN_relation_melting}. Parameters: $W=7\epsilon_0$, $\rho R^2 = 4.4$, $a_{\rm opt} = 1.502$, $(M, L) = (22, 6)$.}
\end{figure}

	The density of thermal excitations $\ev*{\fpsi^\dg(\vb{r})\fpsi(\vb{r})}$ increases with temperature while the condensate wave function is depleted. Typical density profiles are shown in Fig.~\ref{fig:density_profile}. Interestingly, the thermally-excited quasiparticles tend to accumulate on a ring around each of the condensate peaks. Similar density profiles are observed throughout the solid phase of soft-core bosons. 
	The solid phase eventually melts away with increasing temperature. 	Within the HF approximation this occurs as a first order phase transition. The critical temperature is obtained as the crossing between the free energies of the solid phase and the homogeneous phase. At fixed total density $\rho$, it is necessary to use the free energy
	\begin{gather}
		\label{eq:free_energy_def}
		F_{\rm m.f.} = \Omega_{\rm m.f.} + \mu N\,,\\
		\qq*{with} N = \rho A = \int \dd{\vb{r}}\qty\big[\abs{\psi(\vb{r})}^2+\ev*{\fpsi^\dg(\vb{r})\fpsi(\vb{r})}]
	\end{gather} 
	the total particle number, rather than the grand potential $\Omega_{\rm m.f.}$ of Eq.~\eqref{eq:Bogoliubov_inequality}, for comparing different phases. An example of free energy crossing is shown in Fig.~\ref{fig:three_panels}. 
	The critical temperature of the first order phase transition as a function of $W$ is shown as the solid line in the phase diagram (see Fig.~\ref{fig:phase_diagram}). This has been obtained by using, for fixed coupling $W$, the optimal lattice constant of Fig.~\ref{fig:lattice_constant} also at finite temperature. It would be possible to optimize the lattice constant at finite temperature as well by minimizing the free energy with respect to $a$ at fixed density, however this is in practice unnecessary since the lattice constant changes very little with temperature. 

	As shown in Fig.~\ref{fig:density_profile}, the melting of the solid phase occurs for rather small condensate depletion  $\rho_{\rm t}/\rho\sim3\%$, with $\rho_{\rm t} = \rho -\rho_{\rm c} = \int \dd{\vb{r}} \ev*{\fpsi^\dg(\vb{r})\fpsi(\vb{r})}/A$ the average density of the thermal component. Indeed, the density profiles shown in Fig.~\ref{fig:density_profile} are for a point in the phase diagram rather close to the first order phase transition between the density modulated and the homogeneous phase predicted by HF theory. This point is marked in the left panel of Fig.~\ref{fig:phase_diagram}.
	
\subsection{Berezinskii-Kosterlitz-Thouless transitions}
\label{sec:BKT_transitions}

With increasing temperature, the transition from a superfluid to a normal fluid is expected to occur.  Mean-field theory can not describe this transition in 2D. However, the results of HF theory can provide an indirect estimate of the BKT temperature by means of the universal relation with the discontinuity of the superfluid density~\cite{Nelson1977}
\begin{equation}
	\label{eq:KN_relation}
	\frac{\hbar^2\rho_{\rm s}(T_{{\rm BKT, sf}}^-)}{m} = \frac{2k_{\rm B}T_{{\rm BKT, sf}}}{\pi}\,.
\end{equation} 
On the right-hand side, $T^-_{\rm BKT,sf}$ denotes a temperature infinitesimally smaller than the critical temperature $T_{\rm BKT,sf}$ of the superfluid BKT transition.

The superfluid density is defined as the response  of the system to a twist in the boundary conditions. Numerically, it is convenient  to implement twisted boundary conditions by means of the minimal substitution
\begin{equation}
	-i\grad_{\vb{r}} \to -i\grad_{\vb{r}} - \bm{\kappa}\,. 
\end{equation}	
in the kinetic term of the Hamiltonian~\eqref{eq:H_free_first}. Then the superfluid density tensor is defined as
\begin{equation}
	\label{eq:sf_rho_def}
	\frac{\hbar^2}{m}[\rho_{\rm s}]_{i,j} = \frac{1}{A}\pdv[2]{F}{\kappa_i}{\kappa_j}\,,
\end{equation}
in terms of the free energy $F$. Within the HF approximation $F$ is replaced by the mean-field free energy $F_{\rm m.f.}$ given by Eq.~\eqref{eq:free_energy_def}.
Due to the triangular symmetry of the solid lattice, even in the supersolid phase the superfluid density tensor is proportional to the identity, namely $[\rho_{\rm s}]_{i,j} = \rho_{\rm s}\delta_{i,j}$, with the scalar $\rho_{\rm s}$ called the superfluid density in the following.

As shown in Fig.~\ref{fig:three_panels}b, the superfluid density obtained from~\eqref{eq:sf_rho_def} as a function of temperature provides an estimate of $T_{\rm BKT,sf}$ by means of the universal relation~\eqref{eq:KN_relation}. The same procedure is used both in the homogeneous phase and in the solid phase, and the BKT temperatures for the transitions from a superfluid to a normal fluid $T_{\rm BKT,sf}$ and from supersolid to a normal solid $T_{\rm BKT,ssl}$ are shown in Fig.~\ref{fig:phase_diagram} as the dotted and dash-dotted lines, respectively.
Notice that, for the values of $W$ shown in Fig.~\ref{fig:phase_diagram}, the degeneracy temperature at the density $\rho R^2=4.4$, $k_BT^\ast \simeq55\epsilon_0$, is significantly larger than all drawn transition lines. This justifies the use of the HF description in terms of the quasi-condensate density $\rho_c$, whose thermal depletion remains marginal in the whole phase diagram.

Similarly to the superfluid transition, the melting of a quasi-crystal in two dimension has also topological character and can occur through two separate BKT transitions~\cite{Nelson1978,Halperin1978,Nelson1979,Young1979,Gasser2010}. For increasing temperature, the first BKT transition occurs between a solid and the so-called hexatic phase, characterized by short-range positional order and quasi-long range orientational order. The topological excitations are in this case lattice dislocations. A second BKT transition driven by the unbinding of  disclinations leads to the loss of orientational order as well, resulting in a featureless fluid phase at high temperature.

Both melting BKT transitions in two dimensions are characterized by universal relations analogous to Eq.~\eqref{eq:KN_relation}. In the case of the solid-hexatic transition, the universal relation reads~\cite{Halperin1978,Young1979}
\begin{equation}
	\label{eq:KN_relation_melting}
	Y(T_{\rm BKT, hex}^-)a^2 = 16\pi k_{\rm B}T_{\rm BKT, hex}\,.
\end{equation} 
On the left-hand side $Y(T_{\rm BKT, hex}^-)$ is the Young modulus of the two-dimensional quasi-crystal slightly below the BKT temperature $T_{\rm BKT, hex}$ and $a$ is the lattice constant of the triangular lattice. This relation is used here to obtain an estimate of the BKT temperature of the solid-hexatic transition, in the same way as for the superfluid to normal BKT transition.

The Young modulus in two dimensions is related to the  shear $G$ and bulk $B$ moduli by
\begin{equation}
	\label{eq:Young}
	Y = \frac{4}{G^{-1}+B^{-1}}\,.
\end{equation}
To obtain the shear modulus, an area-preserving deformation of the crystalline lattice is imposed, see Appendix~\ref{app:discretization}. In the notation of elasticity theory~\cite{Landau1986}, the corresponding displacement vector field is
\begin{equation}
	\vb{u}(x,y) = \pmqty{y(\cot \theta'-\cot\theta) \\ 0}\,.
\end{equation}
with $\theta$ and $\theta'$ the angles formed by the fundamental vectors~$\vb{a}_1$ and $\vb{a}_2$~\eqref{eq:psi_cond_per}-\eqref{eq:Gamma_cond_per} before and after the deformation, respectively. The only nonzero components of the strain tensor are off-diagonal
\begin{equation}
	\begin{split}
		u_{xy} = u_{yx} = \frac{1}{2}\qty(\pdv{u_x}{y}+\pdv{u_y}{x})  = \frac{\cot \theta'-\cot\theta}{2}\,,
	\end{split}	
\end{equation}
realizing thus a pure shear deformation of the lattice. If the deformation is small, the free energy is a quadratic function of the strain tensor when the lattice is initially in its equilibrium configuration, the triangular lattice ($\theta = 2\pi/3$) in our case. Then, the shear modulus $G$ is obtained as the coefficient of the quadratic term in the free energy variation caused by the lattice deformation
\begin{equation}
	\label{eq:F_theta}
	\begin{split}
	\frac{\Delta F}{A} &\approx G (u_{xy}^2+u_{yx}^2) 
	=  \frac{G}{2}\qty(\cot\theta'+\frac{1}{\sqrt{3}})^2\,.
	\end{split}
\end{equation}
In practice, $G$ is computed through a quadratic fitting of the mean-field free energy as a function of $\theta'$.
Notice that the shear modulus is non-vanishing only in the crystalline phase, while it is zero both in the hexatic and fluid phase. Furthermore, a nonzero shear modulus can be taken as the definition of a solid phase in 2D, since strictly long-range positional order is forbidden due to the Mermin-Wagner theorem.

For the bulk modulus $B$, one considers a displacement field corresponding to a pure hydrostatic compression
\begin{equation}
	\label{eq:hydro_comp}
	\vb{u} = -\lambda\vb{r}\,,\qquad u_{ij} = u_{ii}\delta_{i,j} = -\lambda\delta_{i,j}\,.
\end{equation}
This is done numerically by scaling the fundamental vectors $\vb{a}_i$ by the same factor $(1-\lambda)$, while the area becomes $(1-\lambda)^2A$ (see Appendices~\ref{app:momentum_space}-\ref{app:discretization} and Fig.~\ref{fig:sheared_lattice} for details). 
Then, the bulk modulus is obtained from the free energy variation~\cite{Landau1986}
\begin{equation}
	\label{eq:B_rho_def}
	\begin{split}
	\frac{\Delta F}{A} &\approx  -P(u_{xx}+u_{yy}) +  \frac{B}{2}(u_{xx}+u_{yy})^2 \\&=  2P\lambda +  2B\lambda^2\,.
	\end{split}
\end{equation}
Note that when the above definition is applied, the particle number is kept constant as a function of $\lambda$, thus the first order term is the pressure $P$ and $B$ corresponds to the inverse compressibility of the system. The 2D pressure $P$ is nonzero since the soft-core interaction potential~\eqref{eq:soft_core} is purely repulsive. In contrast, the optimal lattice constant is found by requiring that the first order term in the expansion of the free energy as a function of the lattice vectors length vanishes at fixed average density $\rho = N/A$ (see Fig.~\ref{fig:lattice_constant} and Sec.~\ref{sec:phase_diagram_HF}). In the case of a supersolid, where number of particles incommensurability with respect to the lattice structure is important, it is not obvious whether the quantity $B$ that enters in Eq.~\eqref{eq:Young} should be computed at fixed particle number or particle density~\cite{Rakic2024}. For a discussion of how the BKT temperature $T_{\rm BKT,hex}$ is modified if the latter prescription is used see Appendix~\ref{app:bulk_shear}.

The Young modulus as a function of temperature obtained from the mean-field free energy and the graphical solution of Eq.~\eqref{eq:KN_relation_melting} are exemplified in Fig.~\ref{fig:three_panels}c. The resulting estimate of the BKT temperature $T_{\rm BKT,hex}$ of the transition between the solid phase and the hexatic phase is shown as the dashed line in Fig.~\ref{fig:phase_diagram}.
In drawing the phase diagram, we interpret the temperature $T_{\rm c, sl}$ of the first order transition between the solid and the homogeneous phases predicted by HF theory as the transition line between the hexatic and homogeneous phases. 

    \section{Results of the Path-Integral Monte Carlo simulations}
    \label{sec:PIMC}
    \begin{figure*}
        \centering
	    \includegraphics{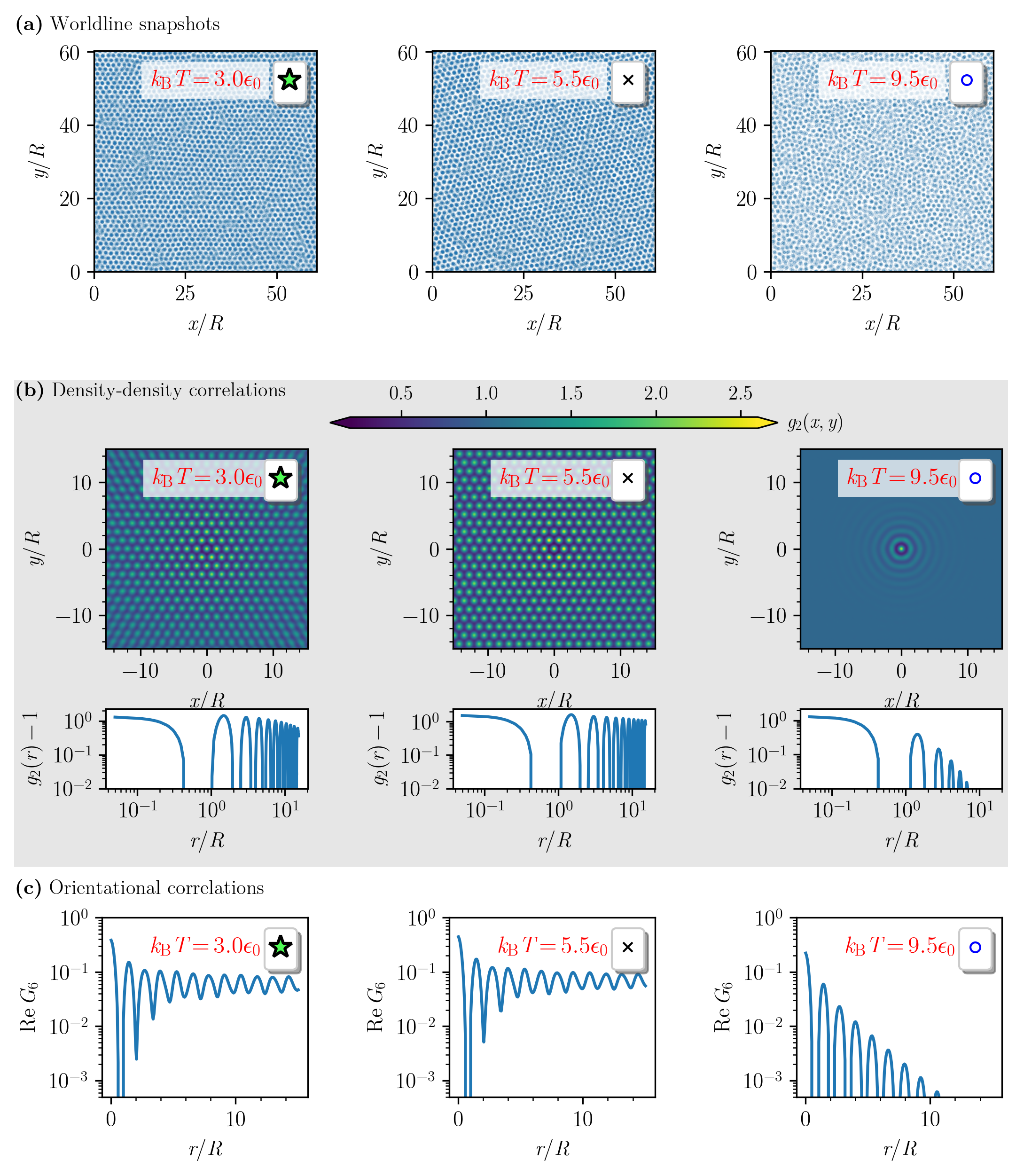}
	    \caption{PIMC simulations for $N=16128$ particles with interaction strength $W=6.6\epsilon_0$ at three values of temperature corresponding to three different phases. Panel (a) shows example snapshots of worldline configurations. Panel (b) shows the density-density correlation functions in the plane $g_2(x,y)$ (upper row) as well as their magnitude along lattice directions (lower row). Panel (c) shows the real part of the orientational correlation functions $G_6(r)$. The symbols displayed in the upper right corner of the plots correspond to the system phase as classified in the phase diagram in Fig.~\ref{fig:phase_diagram}.}\label{fig:pimc_correlations}
    \end{figure*}

    Simulations of the microscopic Hamiltonian with the soft-core interaction potential~\eqref{eq:soft_core} are performed using the continuous-space PIMC algorithm~\cite{Ceperley1995,Boninsegni2006}, in particular with its formulation for the canonical ensemble with periodic boundary conditions described in Ref.~\cite{Spada2022}.
    Our findings, obtained at fixed number of particles $N$, are compared with the low-temperature results of Ref.~\cite{Saccani2011a} obtained at fixed chemical potential, finding agreement.
	To accurately resolve the phase boundaries and capture subtle structural features, we employ the lookup-table technique~\cite{Allen1987}, enabling simulations with a significantly larger number of particles: all the simulations presented in this work are performed with either $N=4032$ or $N=16128$ particles.
    Additional sizes are considered in Appendix~\ref{app:finite_size_pimc} for a more detailed analysis of finite-size effects.
    For these systems we compute, at each point in the parameter space, the relevant order parameters and correlation functions.
	Uncertainties are obtained by statistical average of many independent simulations, started from either a perfect triangular lattice configuration or a homogeneous random configuration and then thermalized until they reached compatible structural and thermodynamical properties.
	Example worldline snapshots for the interaction strength $W=6.6\epsilon_0$ at three temperatures are reported in panel (a) of Fig.~\ref{fig:pimc_correlations}.
	The solid or fluid nature of the system is determined by analyzing the orientational order of neighboring particles~\cite{Nelson1979,Bagchi1996,Bernard2011,Huang2020}. This provides information about the relative positions of the clusters, which, in the solid phase, are known to form a triangular lattice.
	We thus define the six-folded director field
        \begin{equation}
	    \Psi_6(\vec{r}_p) = \frac{1}{N_{p}}\sum_{q} e^{i6\theta_{pq}}\,,
	    \qquad q \mid \vec{r}_q \in \mathcal{D}_{q}\,,
	    \label{eq:psi_6}
        \end{equation}
	where $\theta_{pq}$ is the angle between the vector connecting particles $p$ and $q$ and a reference axis, and $N_p$ is the total number of particles within the disk $\mathcal{D}_{p}$ of radius $2R$ centered at the reference particle position $\vec{r}_p$.
	The above definition of $\Psi_6$ allows one to extract information about the local orientation of neighbouring clusters without the need of identifying and discerning them. The latter task  is in fact not sharply defined.
	The value $2R$ for the disk radius is empirically chosen in such a way that it contains only the nearest clusters, and it coincides with a minimum in the density-density correlation function $g_2(r)$, see panel (b) of Fig.~\ref{fig:pimc_correlations}, and, a posteriori, it also coincides with a minimum in the orientational correlation function $G_6(r)$ defined as
	\begin{equation}
		G_6(r) = \left\langle \Psi_6(\vec{r}_p) \Psi_6^*(\vec{r}_q) \right\rangle \,,
	\end{equation}
	with $r = | \vec{r}_p - \vec{r}_q |$ the distance between two particles.
	We also define the global orientational order parameter, $|\psi_6|^2 = \left\langle | \Psi_6(\vec{r}) |^2 \right\rangle$.
	Systems within the solid phase are characterized by long-ranged orientational correlations (finite $|\psi_6|^2$), while systems within the fluid phase are characterized by short-ranged orientational correlations (vanishing $|\psi_6|^2$).
	An algebraically decaying quasi-long range $G_6(r)$ would signal the hexatic phase but it is difficult to precisely pinpoint with our simulations due to the presence of finite-size effects.
	In the phase diagram of Fig.~\ref{fig:phase_diagram}, points are characterized by (super)solid phases if their $G_6$ correlation functions are well fitted (coefficient of determination $\mathcal{R}^2 > 0.95$) by a damped model $y = A r^{-B} \sin(k r + D ) + C$ with a finite constant $C$.
	Points corresponding to the homogeneous phases are instead determined by fitting the peaks of $G_6$ to an exponentially decaying function $y = A e^{-B r}$ and requiring a coefficient of determination $\mathcal{R}^2 > 0.98$.
	In panel (c) of Fig.~\ref{fig:pimc_correlations} the two behaviors are visible: supersolid and solid phases display $G_6$ oscillating around finite values, while the rightmost plot shows the exponential decay obtained in the normal fluid phase.
	For these points, it is worth analyzing the corresponding density-density correlation functions reported in panel (b): The supersolid and solid phases are characterized by an evident crystalline structure, but with the magnitude of the peaks of $g_2(r) - 1$ decreasing with $r$, in agreement with the Mermin-Wagner theorem. The fluid phase, instead, shows rapidly decaying density-density correlations.
	Points of the phase diagram whose orientational correlation functions don't satisfy any of the two requirements above are labeled as ``unknown'' and may be compatible with an algebraic decay of $G_6(r)$, characteristic of the hexatic phase.
    
	The superfluid nature of the system is determined by comparison of the superfluid fraction of the simulations, computed via the winding number estimator~\cite{Pollock1987,Ceperley1989}, with the universal jump predicted by the Nelson-Kosterlitz equation~\eqref{eq:KN_relation}.
	If the computed superfluid fraction is larger than the universal jump at the transition, the system is superfluid, while, if it is smaller, sizeable finite-size effects are expected to renormalize it to zero and the system is not superfluid.
	When $\rho_s/\rho$ is compatible (within three standard deviations) with the value computed from Eq.~\eqref{eq:KN_relation}, the corresponding points in the phase diagram of Fig.~\ref{fig:phase_diagram} are labeled as unknown.
	This classification clearly separates the diagram in four patches---a homogeneous normal fluid, a homogeneous superfluid, a normal solid, and a supersolid---with boundaries similar to what is predicted by the Hartree-Fock method, although with a sizable rescaling of the temperatures to lower values.
	At the boundary between the solid and homogeneous fluid patches, where the self-consistent Hartree-Fock method predicts the hexatic phase, the PIMC method finds a thin region of points that get classified as unknown, where we observe a slow decay of the $G_6$ correlation function and large finite-size effects.

     \begin{figure*}
        \centering
        \includegraphics{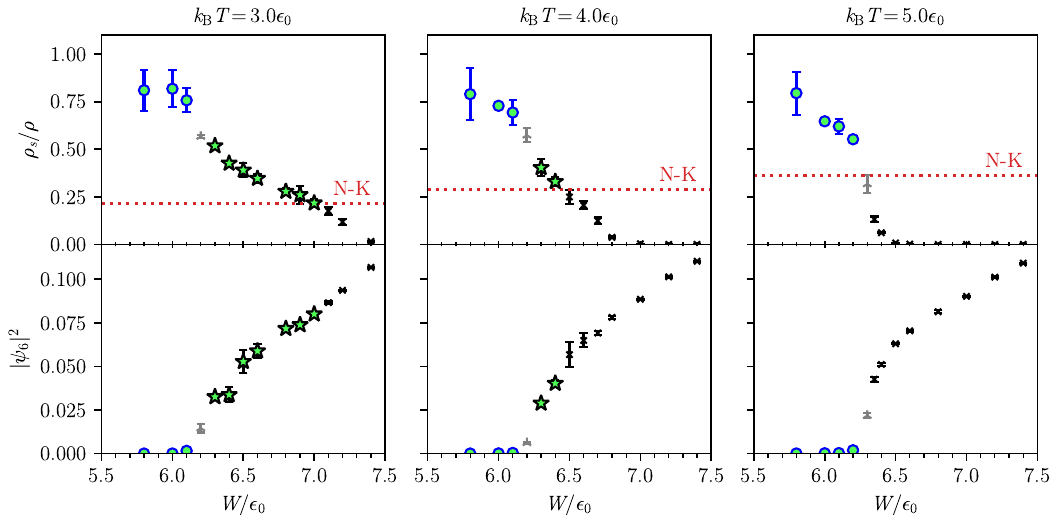}
        \caption{Superfluid fraction $\rho_s/\rho$ and orientational order parameter $|\psi_6|^2$ for simulations with $N=4032$ particles at three different values of temperature.
        Markers are as in Fig.~\ref{fig:phase_diagram}. The horizontal dotted line represents the universal jump of the superfluid density predicted by the Nelson-Kosterlitz equation.}
        \label{fig:pimc_order_param}
    \end{figure*}
	In Fig.~\ref{fig:pimc_order_param} we plot the values of the order parameters as functions of the interaction strength $W$, for three  temperatures $k_B T / \epsilon_0 = 3.0, \, 4.0$, and $5.0$. This allows analyzing how the order parameters change while crossing the different phases.
	At the smallest  interaction strengths $W \lesssim 6.2$, we find a large superfluid fraction and a vanishing orientational order, compatible with the presence of a homogeneous superfluid. As $W$ increases, the superfluid fraction decreases, while $|\psi_6|^2$ becomes finite.
	The rise of the latter order parameter is moderately quick but, due to finite size effects, we cannot conclude whether its appearance would be continuous or discontinuous in the thermodynamic limit. Notice that a discontinuous drop to zero of $|\psi_6|^2$ is compatible with the BKT scenario of the 2D melting transition~\cite{Nelson1979}.
    At the lowest temperatures $k_B T / \epsilon_0 = 3.0$ and $4.0$, the order parameters change less rapidly, almost linearly with $W$. Notably, we find a region with finite superfluid fraction, greater than the universal jump~\eqref{eq:KN_relation}, together with a finite $|\psi_6|^2$, signalling the supersolid. By increasing $W$ even further, the system becomes a normal solid.
	At the highest temperature $5.0 \epsilon_0$, instead, the change in the order parameters is more abrupt, directly taking the system from the homogeneous superfluid phase to the normal solid.
A further analysis on how the finite sizes affect the superfluid and orientational order parameters is provided in Appendix~\ref{app:finite_size_pimc}. It is shown that, at the three temperatures of Fig.~\ref{fig:pimc_order_param}, the superfluid fraction sizably scales with the size only beyond the critical threshold predicted by Eq.~\eqref{eq:KN_relation}. This is consistent with the standard BKT scenario. As already mentioned in Section~\ref{sec:phase-diagram}, a renormalization of the critical threshold is expected close to a quantum critical point~\cite{Doniach1981}, corresponding in this case to the transition from supersolid to normal solid at zero temperature. This effect is not noticeable at the temperatures we address.
    \begin{figure}
        \includegraphics{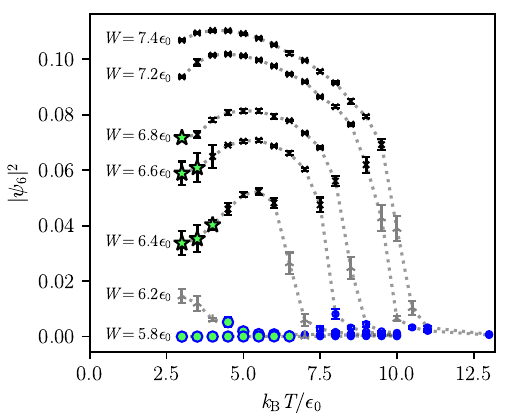}
        \caption{Orientational order parameter $|\psi_6|^2$ for simulations with $N=4032$ particles at fixed values of the interaction $W$.
    Markers are as in fig.~\ref{fig:phase_diagram}.}
        \label{fig:pimc_order_by_disorder}
    \end{figure}

	An interesting feature can be observed by analyzing the behavior of $|\psi_6|^2$ at fixed $W$ as a function of the temperature, shown in Fig.~\ref{fig:pimc_order_by_disorder}.
	At low temperature and intermediate values of $W$, where the superfluid phase is observed, the system increases its orientational order as it is heated, even after the transition to a normal solid.
	After reaching a maximum within the solid phase, thermal fluctuations take over, and the orientational correlations decrease until the solid melts and the system becomes homogeneous with vanishing $|\psi_6|$.
    A similar feature emerges also in the long-range decay of the pair correlation function shown in Fig.~\ref{fig:pimc_correlations} (b) which is more pronounced at the temperature $k_BT=3.0\epsilon_0$ than at $k_BT=5.5\epsilon_0$.
	This peculiar behavior, revealed by extensive simulations in a wide region of the parameter space, is a striking new feature of this system that deviates from the standard phenomenology: an increase of the thermal energy leads to more ordered structures. In fact, orientational order is observed to decrease with increasing temperature in experiments on 2D suspensions of colloidal particles~\cite{Keim2007} and in classical simulations of 2D melting~\cite{Strandburg1988}. This seems to indicate that quantum fluctuations play an important role in the behavior of $|\psi_6|^2$ shown in Fig.~\ref{fig:pimc_order_by_disorder}.

    Furthermore, Fig.~\ref{fig:pimc_order_by_disorder} shows that the appearance of a finite value of $|\psi_6|^2$ occurs quite suddenly as a function of temperature. However, similarly to Fig.~\ref{fig:phase_diagram}, we can not conclude whether the transition is continuous or discontinuous since much larger simulations would be needed to establish precisely the value of $|\psi_6|^2$ for the points marked as unknown in Fig.~\ref{fig:pimc_order_by_disorder}. 

    %
   
\section{Conclusions and discussions}
\label{sec:conclusion}

We have investigated the phase diagram of soft-core bosons as a function of temperature and interaction strength. Special attention was devoted to the interplay between superfluid behavior and quasi-long-range structural order. The latter is a key feature of 2D quantum systems at finite temperatures, requiring specific theoretical treatments. In particular, it is important to address both positional and orientational correlations.  
In detail, we presented analyses based on self-consistent HF calculations and on unbiased PIMC simulations.  
The former method extends earlier zero-temperature studies based on the GP equation and Bogoliubov theory~\cite{Pomeau1994,Macri2013,Rakic2024,Blakie2024a}. The implementation of twisted boundary conditions allowed determining the superfluid response, while the combination of a lattice deformation and a hydrostatic compression allowed the computation of the Young modulus, which provides information on the stability of the quasicrystal against lattice dislocations, possibly leading to a hexatic phase. 
On the other hand, compared to previous PIMC studies, we addressed larger sizes and additional estimators for orientational correlations, allowing us to identify power-law decays of correlations and to discern quasicrystal, fluid, and potentially hexatic phases. The standard winding number estimator was used to characterize the superfluid behavior.

The analysis of phase diagram revealed the occurrence of the following phases: normal and superfluid homogeneous phases, a normal quasicrystal, a superfluid quasicrystal (i.e. the supersolid), and possibly a hexatic phase. According to HF theory, the latter phase is stable in a sizable window  separating the normal fluid from the solid. In PIMC simulations, this window shrinks and possibly vanishes, but an unambiguous assessment is not possible due to the finite system sizes. 
The phase boundaries predicted by the HF theory turn out to be in qualitative agreement with the PIMC results. This finding establishes the former method as a useful computational tool for the rapid exploration of phase diagrams of novel physical systems. 
On the other hand, PIMC simulations equipped with estimators for positional and orientational orders allowed us to inspect the quasi-long range nature of off-diagonal and structural correlations that characterize 2D systems. Interestingly, we identified a parameter regime where the orientational order increases with the temperature, which represents an instance of the order-by-disorder mechanism. 
Future investigations should focus on different interaction potentials, trying to identify promising candidates for the observations of hexatic phases in the quantum realm, eventually also demonstrating superfluid properties. 
On the technical side, it would be interesting to develop a stable numerical algorithm for solving the self-consistent equations of HFB theory, see Sec.~\ref{sec:HF_variational}, that should provide a quantitatively more accurate phase diagram compared to HF theory especially a low temperature, where quantum fluctuations dominate over thermal fluctuations.  
   
\begin{acknowledgments}

We acknowledge support from: 
PNRR MUR project PE0000023-NQSTI; 
PRIN 2022 MUR project ``Hybrid algorithms for quantum simulators'' -- 2022H77XB7; 
PRIN-PNRR 2022 MUR project ``UEFA'' -- P2022NMBAJ;
CINECA awards IsCc2\_REASON, IsCc2\_QD2DBS, and INF24\_lincoln;
EuroHPC Joint Undertaking for awarding access to the EuroHPC supercomputer LUMI, hosted by CSC (Finland), through EuroHPC Development and Regular Access calls;
ICSC - Centro Nazionale di Ricerca in HPC, Big Data and Quantum Computing, CN00000013 Spoke 7 -- Materials \& Molecular Sciences, funded by the European Union under NextGenerationEU;
Research Council of Finland under Grants No. 330384, No. 336369, and No. 358150; 
Aalto Science-IT project for the computational resources;
Provincia Autonoma di Trento. 
S. Peotta gratefully acknowledges the support and kind hospitality of the Pitaevskii Center for Bose-Einstein Condensation (Trento, Italy) and its members during a yearlong visit in 2023–2024. A substantial part of the work leading to this publication was carried out during this research visit.
   	\end{acknowledgments}

	\appendix
	
	\section{Self-consistency equations}
	\label{app:self-consistent}
	
	The purpose of this Appendix is to provide some details regarding the derivation of the extended GP equation~\eqref{eq:extGP_cont_1} and self-consistency equation for the HF potential~\eqref{eq:sc_cont_1}-\eqref{eq:sc_cont_3}. Here, the same approach as developed in~\cite{Peotta2022,Tam2024} in the fermionic case is used. It is based on the Legendre transform that allows to switch from the variational parameters $\xi(\vb{r}),\,\Gamma(\vb{r}, \vb{r}')$ to the corresponding expectation values $\ev*{\hat{\psi}(\vb{r})} = \psi(\vb{r}),\,\ev*{\hat{\psi}^\dg(\vb{r})\hat{\psi}(\vb{r}')}$ as the independent variables in the mean-field grand potential. The essential new element in the bosonic case, with no counterpart for fermions, is the presence of the linear term in $\ham_0$~\eqref{eq:H0} associated to the field $\xi(\vb{r})$, which leads to a finite expectation value of the field operators, that is to the condensate wave function~$\psi(\vb{r})$. How this can be taken into account is explained in detail below.
	
	As a first step, note that the expectation values of the field operators and their bilinear combinations are obtained by taking the functional derivatives of the auxiliary free energy $\Omega_0$ with respect to the variational parameters
	\begin{gather}
		\label{eq:psi_def}
		\fdv{\Omega_0}{\xi^*(\vb{r})} = \psi(\vb{r})\,,\quad 	
		\fdv{\Omega_0}{\xi(\vb{r})} = \psi^*(\vb{r})\,,\\
		\label{eq:ev_bilinear}
		\fdv{\Omega_0}{\Gamma(\vb{r},\vb{r}')} = \ev*{\hat{\psi}^\dg(\vb{r})\hat{\psi}(\vb{r}')}\,.
	\end{gather}
	Then, Wick's theorem is used to compute the expectation value of the interaction (quartic) term $\ev*{\ham_{\rm int}}$ in the Hamiltonian $\ham$. Note that Wick's theorem can not be applied directly since the variational Hamiltonian $\ham_0$ is not purely quadratic, but contains a term linear in the field operators. On the other hand, using the standard algebra of coherent states and displacement operators~\cite{Puri2001}, one can show that the expectation values of product of fluctuation operators $\fpsi(\vb{r})$ and $\fpsi^\dg(\vb{r})$~\eqref{eq:fluc_op} do not depend on the field $\xi(\vb{r})$. Moreover, the following relation between the field $\xi(\vb{r})$ and the condensate wave function $\psi(\vb{r})$ holds
	\begin{gather}
		\xi = -H_0 \psi \,, \label{eq:xi_H0_psi}\\
		\label{eq:BdG_Ham}
		 H_0 = H_{\rm free} -\mu + \Gamma =
		-\frac{\hbar^2\nabla^2}{2m} + U -\mu +\Gamma\,.
	\end{gather}
	Here, $H_0$ is the analogue of the Bogoliubov-de Gennes Hamiltonian in fermionic systems, that is the single-particle operator that gives the quasiparticle excitations. In order to derive~\eqref{eq:xi_H0_psi}, it is necessary for $H_0$ to be invertible. In addition, for bosonic systems, $H_0$ is required to be a positive semidefinite operator (no negative eigenvalues). In general, it is found that for a self-consistent solution of the mean-field problem the operator $H_0$ has a finite gap (the lowest eigenvalue is strictly positive) and thus it is positive definite and invertible, as required for~\eqref{eq:xi_H0_psi} to hold.
 	
	According to the above discussion, the expectation values of products of fluctuation operators can be computed by setting $\xi(\vb{r}) = 0$, thus the variational Hamiltonian $\ham_0$ is purely quadratic and Wick's theorem can be applied. As a consequence, to compute expectation values of arbitrary products of field operators, such as $\ev*{\hat{\psi}^\dg(\vb{r}_1)\hat{\psi}^\dg(\vb{r}_2) \hat{\psi}(\vb{r}_3)\hat{\psi}(\vb{r}_4)}$, one has to first use~\eqref{eq:fluc_op} in the form $\hat\psi(\vb{r}) = \psi(\vb{r}) + \fpsi(\vb{r})$ to replace the field operators with the fluctuation operators and then to apply Wick's theorem to express everything in terms of expectation values of products of two fluctuation operators. Expectation values of an odd number of fluctuation operators always vanish. One obtains for instance
	\begin{equation}
		\label{eq:generalized_Wick_theo}
		\begin{split}
		&\ev*{\hat{\psi}^\dg(\vb{r}_1)\hat{\psi}^\dg(\vb{r}_2) \hat{\psi}(\vb{r}_3)\hat{\psi}(\vb{r}_4)} \\ &\quad =   \ev*{\hat{\psi}^\dg(\vb{r}_1)\hat{\psi}(\vb{r}_4)}\ev*{\hat{\psi}^\dg(\vb{r}_2) \hat{\psi}(\vb{r}_3)} \\ 
		&\quad\quad+\ev*{\hat{\psi}^\dg(\vb{r}_1)\hat{\psi}(\vb{r}_3)}\ev*{\hat{\psi}^\dg(\vb{r}_2) \hat{\psi}(\vb{r}_4)} \\
		&\quad\quad - \psi^*(\vb{r}_1)\psi^*(\vb{r}_2)\psi(\vb{r}_3)\psi(\vb{r}_4)\,.
		\end{split}
	\end{equation} 
	 Without the last term, this reduces to Wick's theorem for bosonic operators in the number conserving case. The condensate wave function, which is nonzero if the field $\xi(\vb{r})$ is nonzero, enters both in the last term, not included in the standard Wick's theorem, and also in the expectation value of field operator bilinears according to~\eqref{eq:fluc_prop_2}. 
	
	Using~\eqref{eq:generalized_Wick_theo}, one can compute the expectation value of the interaction term in the many-body Hamiltonian
	\begin{equation}
		\label{eq:ev_H_int}
		\begin{split}
			\ev*{\ham_{\rm int}} = 
			\frac{1}{2}&\iint\dd{\vb{r}}\dd{\vb{r}'}V(\vb{r}-\vb{r}')	\\
			&\times \big[\ev*{\hat{\psi}^\dg(\vb{r})\hat{\psi}(\vb{r})}\ev*{\hat{\psi}^\dg(\vb{r}') \hat{\psi}(\vb{r}')}
			\\
			& \hspace{0.7cm} + \ev*{\hat{\psi}^\dg(\vb{r})\hat{\psi}(\vb{r}')}\ev*{\hat{\psi}^\dg(\vb{r}') \hat{\psi}(\vb{r})} \\
			 &  \hspace{0.7cm}-\abs{\psi(\vb{r})}^2\abs{\psi(\vb{r}')}^2\big]\,.
		\end{split}
	\end{equation} 
	This last result is useful since the expectation value of the interaction term has been written as an explicit function of the expectation values of the field operator bilinears and the condensate wave function.
	
	After subtracting the expectation value of the interaction term $\ev*{\ham_{\rm int}}$ from the mean-field grand potential $\Omega_{\rm m.f.}$, one is left with the quantity
	\begin{equation}
		\begin{split}
		&\overline{\Omega}_0 = \Omega_0 - \int \dd{\vb{r}} \qty\big[\xi(\vb{r})\psi^*(\vb{r})+\xi^*(\vb{r})\psi(\vb{r})] \\
		&\hspace{1.5cm}-\iint \dd{\vb{r}} \dd{\vb{r}'} \Gamma(\vb{r},\vb{r}')\ev*{\hat{\psi}^\dg(\vb{r})\hat{\psi}(\vb{r}')}\,.
		\end{split}
 	\end{equation}
	It follows from~\eqref{eq:psi_def}-\eqref{eq:ev_bilinear} that $\psi(\vb{r})$ and $\ev*{\hat{\psi}^\dg(\vb{r})\hat{\psi}(\vb{r}')}$, are the conjugate variables of $\xi^*(\vb{r})$ and $\Gamma(\vb{r},\vb{r}')$ respectively, thus one can identify $\overline{\Omega}_0$ as the functional Legendre transform of $\Omega_0$. The Legendre transform $\overline{\Omega}_0$ is a naturally a function of the conjugate variables and its functional derivatives are given by
	\begin{gather}
		\label{eq:cv_der_1}
		\fdv{\overline{\Omega}_0}{\psi(\vb{r})} = -\xi^*(\vb{r})\,,\quad 	\fdv{\overline{\Omega}_0}{\psi^*(\vb{r})} = -\xi(\vb{r})\,,\\
		\label{eq:cv_der_2}
		\fdv{\overline{\Omega}_0}{\ev*{\hat{\psi}^\dg(\vb{r})\hat{\psi}(\vb{r}')}} = -\Gamma(\vb{r},\vb{r}')\,.
	\end{gather} 
	Then, the mean-field grand potential can be written as
	\begin{equation}
		\Omega_{\rm m.f.} = \overline{\Omega}_0 + \ev*{\ham_{\rm int}}\,,
	\end{equation}  
	and its functional derivatives with respect to the conjugate variables can be easily computed from~\eqref{eq:cv_der_1}-\eqref{eq:cv_der_2} and~\eqref{eq:ev_H_int}. One has to keep in mind that, when taking the derivatives, the conjugate variables, namely the fields $\psi(\vb{r})$, $\psi^*(\vb{r})$, $\ev*{\hat\psi^\dg (\vb{r})\hat\psi(\vb{r}')}$, are considered as independent variables, regardless of the relation in~\eqref{eq:fluc_prop_2}.
	Then, requiring that the functional derivatives vanish leads immediately to the self-consistency equation~\eqref{eq:sc_cont_1}-\eqref{eq:sc_cont_3} and to the following relation between the field $\xi(\vb{r})$ and the condensate wavefunction 
	\begin{gather}
		\label{eq:self_cons_psi}
		\xi(\vb{r}) = -\int\dd{\vb{r}'}V(\vb{r}-\vb{r}')\abs{\psi(\vb{r}')}^2\psi(\vb{r})\,.
	\end{gather}
	The extended GP equation is obtained by eliminating the field $\xi(\vb{r})$ in the above equations using~\eqref{eq:xi_H0_psi} and inserting the self-consistency equation~\eqref{eq:sc_cont_1}-\eqref{eq:sc_cont_3} in place of the HF potential. Finally, one recovers the GP equation in the form~\eqref{eq:extGP_cont_1} by replacing the field operators with the fluctuation operators using~\eqref{eq:fluc_prop_2}. It is clear from this derivation that the extended GP equation is equivalent to the self-consistency equation for the variational field $\xi(\vb{r})$~\eqref{eq:self_cons_psi}.

	\section{Momentum-space representation and discrete translational symmetry}
	\label{app:momentum_space}
	
	The numerical solution of the mean-field problem, that is the problem of finding $\psi(\vb{r})$ and $\Gamma(\vb{r},\vb{r}')$ satisfying the extended GP equation~\eqref{eq:extGP_cont_1} and the self-consistency equation~\eqref{eq:sc_cont_1}-\eqref{eq:sc_cont_3}, becomes less computationally demanding by taking advantage of the symmetries of the system, in particular translational symmetry. The Hamiltonian $\ham$ possesses continuous translational symmetry if the single-particle potential $U(\vb{r})$ is a constant. On the other hand, continuous translational symmetry is spontaneously broken down to discrete translational symmetry in the solid phase. The purpose of this Appendix is to recast the self-consistency equations, which include the GP equation, in a momentum space representation that allows to take advantage of discrete translational symmetry.  
	
	For the numerical solution, a finite system (also called supercell) with period boundary conditions is considered. The shape and size of the supercell are specified by a basis of vectors $\vb{R}_i$ with $i = 1,2$ for a two-dimensional system, thus the area of the supercell is $A=\abs{\vb{R}_1\times \vb{R}_2}$.
	Periodic boundary conditions imply that
	\begin{gather}
		\label{eq:period_psi}
		\psi(\vb{r}) = \psi(\vb{r}+\vb{R}_i)\,,\\
		\label{eq:period_Gamma}
		\Gamma(\vb{r},\vb{r}') = \Gamma(\vb{r}+\vb{R}_i,\vb{r}') = \Gamma(\vb{r},\vb{r}'+\vb{R}_i)\,.  
	\end{gather}  
	The field $\xi(\vb{r})$ in~\eqref{eq:H0} has been eliminated in favor of the condensate wave function, therefore it is set to zero everywhere in the following.
	
	The periodicity condition~\eqref{eq:period_Gamma} implies that 
	the HF potential can be expanded as a Fourier series. For a function $f(\vb{r},\vb{r}')$ of two arguments such as $\Gamma(\vb{r},\vb{r}')$ the following convention for the Fourier series is used
	\begin{gather}
		\label{eq:Fourier_convention_1}
		f(\vb{r}, \vb{r}') = \frac{1}{A}\sum_{\vb{k},\vb{k}'} f_{\vb{k},\vb{k}'}e^{i(\vb{k}\cdot \vb{r}-\vb{k}'\cdot \vb{r}')}\,,\\
		\label{eq:Fourier_convention_2}
		f_{\vb{k},\vb{k}'} = \frac{1}{A}\iint \dd{\vb{r}} \dd{\vb{r}'} f(\vb{r}, \vb{r}') e^{-i(\vb{k}\cdot \vb{r}-\vb{k}'\cdot \vb{r}')}\,,
	\end{gather}
	where the sum is over all wave vectors $\vb{k}$, $\vb{k}'$ such that 
	\begin{equation}
		\label{eq:kR_cond}
		\vb{k} \cdot \vb{R}_j = 2\pi m_j \qq{with $m_j$ integer.} 
	\end{equation}
	Note that the single-particle potential $U(\vb{r})$ should be considered a function of two arguments, since
	\begin{equation}
		\begin{split}
			&\int \dd{\vb{r}}	\hat{\psi}^\dg(\vb{r}) U(\vb{r})\hat{\psi}(\vb{r}) \\
			&= \iint \dd{\vb{r}}\dd{\vb{r}'}	\hat{\psi}^\dg(\vb{r}) U(\vb{r})\delta(\vb{r}-\vb{r}')\hat{\psi}(\vb{r}')\,.
		\end{split}
	\end{equation}
	Thus, we have for its Fourier transform 
	\begin{equation}
		\begin{split}
			U_{\vb{k},\vb{k}'} &= \frac{1}{A} \iint \dd{\vb{r}} \dd{\vb{r}'} U(\vb{r})\delta(\vb{r}- \vb{r}') e^{-i(\vb{k}\cdot \vb{r}-\vb{k}'\cdot \vb{r}')} \\
			&= \frac{1}{A} \int \dd{\vb{r}} U(\vb{r}) e^{-i(\vb{k} -\vb{k}')\cdot \vb{r}}\,.
		\end{split}
	\end{equation}
	It is clear that the Fourier coefficients depend only in the difference $\vb{k}-\vb{k}'$, therefore we define $U_{\vb{k}-\vb{k}'}= U_{\vb{k},\vb{k}'}$. The external potential $U(\vb{r})$ serves as a source term to induce a translational symmetry breaking pattern with a specific periodicity and phase.
	
	For a function of a single argument, the following Fourier series convention
	is adopted
	\begin{gather}
		\label{eq:psi_Four_conv_1}
		\psi(\vb{r}) = \frac{1}{\sqrt{A}} \sum_{\vb{k}} \psi_{\vb{k}}e^{i\vb{k}\cdot \vb{r}}\,,\\
		\label{eq:psi_Four_conv_2}
		\psi_{\vb{k}} = \frac{1}{\sqrt{A}}\int \dd{\vb{r}} \psi(\vb{r}) e^{-i\vb{k}\cdot \vb{r}}\,.
	\end{gather}
	As before, the sum runs over wave vectors which satisfy the condition~\eqref{eq:kR_cond}. Similarly, for the field operator, we have the expansion
	\begin{gather}
		\label{eq:field_Four_conv_1}
		\hat{\psi}(\vb{r}) = \frac{1}{\sqrt{A}} \sum_{\vb{k}} \hat{b}_{\vb{k}}e^{i\vb{k}\cdot \vb{r}}\,,\\
		\label{eq:field_Four_conv_2}
		\hat{b}_{\vb{k}} = \frac{1}{\sqrt{A}}\int \dd{\vb{r}} \hat{\psi}(\vb{r}) e^{-i\vb{k}\cdot \vb{r}}\,.		
	\end{gather}
	The field operators in momentum space $\hat{b}_{\vb{k}}$ satisfy the canonical commutation relations $[\hat{b}_{\vb{k}},\hat{b}_{\vb{k}'}^\dg] = \delta_{\vb{k},\vb{k}'}$.
	
	Using the conventions just established the variational Hamiltonian reads
	\begin{gather}
		\label{eq:H0_k_space}
		\ham_{0} = \sum_{\vb{k},\vb{k}'}\hat{b}_{\vb{k}}^\dg[H_0]_{\vb{k},\vb{k}'}\hat{b}_{\vb{k}'} \,,\\
		\label{eq:H0_k_space_single}
		[H_0]_{\vb{k},\vb{k}'} = \qty(\frac{\hbar^2\vb{k}^2}{2m}-\mu)\delta_{\vb{k},\vb{k}'}+U_{\vb{k}-\vb{k}'}+\Gamma_{\vb{k},\vb{k}'}\,.
	\end{gather}
	Here $[H_0]_{\vb{k},\vb{k}'}$ denotes the matrix elements in the momentum space representation of the quasiparticle Hamiltonian $H_0$, which has been introduced previously in~\eqref{eq:BdG_Ham}.   
	The Hermiticity of $H_0$ requires that $\Gamma_{\vb{k},\vb{k}'} = \Gamma_{\vb{k}',\vb{k}}^*$ and $U_{\vb{k}} = U_{-\vb{k}}^*$.

	The momentum space version of the self-consistency equation for the HF potential~\eqref{eq:sc_cont_1}-\eqref{eq:sc_cont_3} reads
	\begin{gather}
		\label{eq:sc_cont_1_ksp}
		\Gamma_{\vb{k},\vb{k}'} = \Gamma^{\rm H}_{\vb{k},\vb{k}'} + \Gamma^{\rm F}_{\vb{k},\vb{k}'}\,,
		\\
		\label{eq:sc_cont_2_ksp}
		\Gamma^{\rm H}_{\vb{k},\vb{k}'} = \frac{1}{A}
		\sum_{\vb{q}}V_{\vb{k}-\vb{k}'}\qty\big(\psi_{\vb{q}}^*\psi_{\vb{q}+\vb{k}-\vb{k}'}+\ev*{\hat{b}_{\vb{q}}^\dg\hat{b}_{\vb{q}+\vb{k}-\vb{k}'}})\,,
		\\
		\label{eq:sc_cont_3_ksp}
		\Gamma^{\rm F}_{\vb{k},\vb{k}'} = \frac{1}{A} \sum_{\vb{q}}V_{\vb{q}}\qty\big(\psi^*_{\vb{k}'+\vb{q}}\psi_{\vb{k}+\vb{q}}+\ev*{\hat{b}^\dg_{\vb{k}'+\vb{q}}\hat{b}_{\vb{k}+\vb{q}}})\,,\\
		\label{eq:Fourier_V}
		\qq*{with} V_{\vb{k}} = \int \dd{\vb{r}} V(\vb{r})e^{-i\vb{k}\cdot \vb{r}}\,.
	\end{gather}
	The interaction potential  $V(\vb{r}) = V^*(\vb{r})$ is a real function, thus its Fourier transform has the property $V_{\vb{k}} = V_{-\vb{k}}^*$, while the symmetry property $V(\vb{r}) = V(-\vb{r})$ translates into $V_{\vb{k}} = V_{-\vb{k}}$. 
	Recall that all the expectation values of the field operators $\hat{b}_{\vb{k}}$ and $\hat{b}_{\vb{k}'}^\dg$ on the right-hand side of the above equations are evaluated with respect to the quadratic Hamiltonian in~\eqref{eq:H0_k_space}, which does not contain any linear term. Thus,  the expectation values in~\eqref{eq:sc_cont_2_ksp}-\eqref{eq:sc_cont_3_ksp} are in fact expectation values of fluctuation operators.
	In the same way as the single-particle potential $U(\vb{r})$, the Hartree potential is local and depends only on the difference of wave vectors, namely $\Gamma^{\rm H}_{\vb{k}-\vb{k}'} =\Gamma^{\rm H}_{\vb{k},\vb{k}'}$, as shown in~\eqref{eq:sc_cont_2_ksp}.
	
	The numerical solution of the extended GP equations is performed in practice by minimizing the functional
	\begin{equation}
		\label{eq:GP_functional}
		\begin{split}
			&F_{\rm eGP}[\psi(\vb{r}), \psi^*(\vb{r})] \\ &= \int \dd{\vb{r}} \psi^*(\vb{r})\qty(-\frac{\hbar^2 \nabla^2_{\vb{r}}}{2m}+U(\vb{r}))\psi(\vb{r})  \\
			&+ \int\int \dd{\vb{r}}\dd{\vb{r}'}V(\vb{r}-\vb{r}')\bigg[\frac{1}{2}\abs{\psi(\vb{r})}^2\abs{\psi(\vb{r}')}^2 \\
			&+\abs{\psi(\vb{r})}^2\ev*{\widetilde{\psi}^\dg(\vb{r}')\widetilde{\psi}(\vb{r}')} + \psi^*(\vb{r})\psi(\vb{r}')\ev*{\widetilde{\psi}^\dg(\vb{r}')\widetilde{\psi}(\vb{r})}\bigg]\,,
		\end{split}
	\end{equation}
	which is the part of the mean-field free energy $F_{\rm m.f.}$~\eqref{eq:free_energy_def} that depends on the condensate wave function. The above functional can be written in momentum space as
	\begin{gather}
		\label{eq:F_eGP_k_space}
		\begin{split}
			&F_{\rm eGP}[\psi_{\vb{k}}, \psi^*_{\vb{k}}] 
			= \sum_{\vb{k},\vb{k}'} \psi^*_{\vb{k}} K_{\vb{k},\vb{k}'}
			\psi_{\vb{k}'} \\
			&\hspace{1cm}+ \frac{1}{2A} \sum_{\vb{k},\vb{k}',\vb{q}}V_{\vb{q}} \psi_{\vb{k}}^*\psi_{\vb{k}+\vb{q}} \psi^*_{\vb{k}'}\psi_{\vb{k}'-\vb{q}}\,,
		\end{split}
	\end{gather}
	where the matrix elements of $K$ are given by
	\begin{gather}
		\label{eq:Akk_def}
		\begin{split}
			&K_{\vb{k},\vb{k}'} = K^*_{\vb{k}',\vb{k}} = \frac{\hbar^2\vb{k}^2}{2m}\delta_{\vb{k},\vb{k}'}+U_{\vb{k}-\vb{k}'} \\
			&+ \frac{1}{A} \sum_{\vb{q}} \qty(V_{\vb{k}-\vb{k}'} \ev*{\hat{b}^\dg_{\vb{q}} \hat{b}_{\vb{q}+\vb{k}-\vb{k}'}} 
			+ V_{\vb{q}}\ev*{\hat{b}^\dg_{\vb{k}'+\vb{q}}\hat{b}_{\vb{k}+\vb{q}}})\,,
		\end{split}
	\end{gather}
	The extended GP equation in momentum space is obtained from the gradient of~\eqref{eq:F_eGP_k_space}
	\begin{equation}
		\begin{split}
			\pdv{F_{\rm eGP}}{\psi_{\vb{k}}^*} 
			&= 
			\sum_{\vb{k}'}K_{\vb{k},\vb{k}'} \psi_{\vb{k}'} 
			+\frac{1}{A}\sum_{\vb{k}',\vb{q}}V_{\vb{q}}\psi_{\vb{k}'}^*\psi_{\vb{k}'-\vb{q}}\psi_{\vb{k}+\vb{q}} \,.
		\end{split}
	\end{equation}
	For the purpose of numerical minimization, it is also useful to compute the Hessian of the GP functional, 
	\begin{equation}
		\label{eq:Hessian}
		\begin{split}
			&\pmqty{\pdv{F_{\rm eGP}}{\psi_{\vb{k}}^*}{\psi_{\vb{k}'}} & \pdv{F_{\rm eGP}}{\psi_{\vb{k}}^*}{\psi_{\vb{k}'}^*}\\[2mm]
				\pdv{F_{\rm eGP}}{\psi_{\vb{k}}}{\psi_{\vb{k}'}} & \pdv{F_{\rm eGP}}{\psi_{\vb{k}}}{\psi^*_{\vb{k}'}} } = 
			\pmqty{K_{\vb{k},\vb{k}'} & 0 \\ 0 & K_{\vb{k},\vb{k}'}^* } \\
			&\hspace{0.6cm}+ \frac{1}{A}\sum_{\vb{q}}V_{\vb{k}-\vb{k}'}\pmqty{\psi_{\vb{q}}^*\psi_{\vb{q}+\vb{k}-\vb{k}'} &  0 \\ 0 & \psi_{\vb{q}+\vb{k}-\vb{k}'}^*\psi_{\vb{q}}} 
			\\
			&\hspace{0.6cm}+\frac{1}{A}\sum_{\vb{q}}V_{\vb{q}}
			\pmqty{\psi^*_{\vb{k}'+\vb{q}}\psi_{\vb{k}+\vb{q}} & \psi_{\vb{k}'-\vb{q}}\psi_{\vb{k}+\vb{q}} \\
				\psi^*_{\vb{k}'-\vb{q}}\psi^*_{\vb{k}+\vb{q}} & 
				\psi_{\vb{k}'+\vb{q}}\psi^*_{\vb{k}+\vb{q}} }\,.
		\end{split}
	\end{equation}
	When the self-consistency equations~\eqref{eq:sc_cont_1_ksp}-\eqref{eq:sc_cont_3_ksp} are satisfied, the Hessian  is related to the quasiparticle Hamiltonian $H_0$ introduced in~\eqref{eq:H0_k_space}. Indeed,  it is straightforward to show that
	\begin{gather}
		\label{eq:Hessian_2}
		\begin{split}
			\pmqty{\pdv{F_{\rm eGP}}{\psi_{\vb{k}}^*}{\psi_{\vb{k}'}} & \pdv{F_{\rm eGP}}{\psi_{\vb{k}}^*}{\psi_{\vb{k}'}^*}\\[2mm]
				\pdv{F_{\rm eGP}}{\psi_{\vb{k}}}{\psi_{\vb{k}'}} & \pdv{F_{\rm eGP}}{\psi_{\vb{k}}}{\psi^*_{\vb{k}'}} } &= 
			\pmqty{[H_0]_{\vb{k},\vb{k}'} & L_{\vb{k},\vb{k}'} \\ L_{\vb{k},\vb{k}'}^* & [H_0]_{\vb{k},\vb{k}'}^* }\,,
		\end{split}
	\\
	\qq*{with} L_{\vb{k},\vb{k}'} = \frac{1}{A}\sum_{\vb{q}}  V_{\vb{q}} \psi_{\vb{k}'-\vb{q}}\psi_{\vb{k}+\vb{q}}\,.
	\end{gather}

	When continuous translational symmetry is preserved, the condensate wave function is a simple plane wave
	\begin{equation}
		\label{eq:psi_c_kc}
		\psi(\vb{r}) = \psi_{\rm c}e^{i\vb{k}_{\rm c}\cdot \vb{r}}\,,
	\end{equation}
	and the HF potential depends only on the difference of the position vectors
	\begin{gather}
		\label{eq:Gamma_tr_inv}
		\Gamma(\vb{r},\vb{r}') = \Gamma(\vb{r}-\vb{r}')\,.
	\end{gather}
	This implies that the Fourier coefficients have the form
	\begin{gather}
		\label{eq:Gamma_k_tr_inv}
		\Gamma_{\vb{k},\vb{k}'} = \delta_{\vb{k},\vb{k}'}\Gamma_{\vb{k}}\qq{with} \Gamma_{\vb{k}} = \int \dd{\vb{r}} \Gamma(\vb{r})e^{-i\vb{k}\cdot \vb{r}}\,.
	\end{gather}
	Furthermore, if time-reversal symmetry is not broken, one can set $\vb{k}_{\rm c} = 0$. 
	
	In the case of spontaneous breaking of translational invariance, the condensate wave function and the HF potential have a periodicity characterized by the fundamental vectors $\vb{a}_{i=1,2}$ according to~\eqref{eq:psi_cond_per}-\eqref{eq:Gamma_cond_per}.
	In analogy with~\eqref{eq:psi_c_kc}, it would be possible to assume a more general condition than~\eqref{eq:psi_cond_per}, namely that the condensate wave function is a Bloch plane wave with arbitrary quasimomentum $\vb{k}_{\rm c}$ [$\psi(\vb{r}+ \vb{a}_i) = e^{i\vb{k}_{\rm c}\cdot \vb{r}}\psi(\vb{r})$]. However, time-reversal symmetry is not expected to be broken in the supersolid phase, therefore only the case in which~\eqref{eq:psi_cond_per} holds is considered in the present work.   
	
	The symmetry breaking pattern specified by the constraints~\eqref{eq:psi_cond_per}-\eqref{eq:Gamma_cond_per} must be compatible with periodic boundary conditions. One possible way to ensure compatibility (although not the most general one) is to require
	\begin{equation}
		\label{eq:R_rel_a}
		\vb{R}_i = L_i\vb{a}_i\qq{with} i = 1,2\,.
	\end{equation}
	The positive integers $L_i$ specify the number of unit cells associated to  crystalline order
	along the respective directions $\vb{R}_i$.
	
	The periodicity constraints~\eqref{eq:psi_cond_per}-\eqref{eq:Gamma_cond_per} enforce some selection rules on the Fourier coefficients in the expansions~\eqref{eq:psi_Four_conv_1} and~\eqref{eq:Fourier_convention_1}. These are best expressed by means of the reciprocal lattice obtained from the basis vectors $\vb{a}_i$, denoted as
	\begin{equation}
		\label{eq:rec_lattice_def}
		\mathcal{R}_{\vb{a}_i} = \big\{\vb{g}\,\,  \big| \,\,\vb{g}\cdot\vb{a}_i = 2\pi n_i\qq{with} n_i \in \mathbb{Z}  \big\}\,.
	\end{equation} 
	The reciprocal lattice is generated by taking integer linear combinations of the basis vectors $\vb{g}_i$, defined by $\vb{g}_i\cdot \vb{a}_j = 2\pi\delta_{i,j}$. In the case of the Fourier coefficients of the condensate wave function, the selection rule is
	\begin{equation}
		\label{eq:psi_sel_rule}
		\psi_{\vb{k}} \neq  0 \Rightarrow \vb{k} \in \mathcal{R}_{\vb{a}_i}\,.
	\end{equation}
	For the Fourier coefficients of the HF potential one has
	\begin{equation}
		\label{eq:f_sel_rule}
		\Gamma_{\vb{k},\vb{k}'} \neq 0 \Rightarrow \vb{k}-\vb{k}' \in \mathcal{R}_{\vb{a}_i}\,.
	\end{equation}
	In other words, the nonzero coefficients can be parametrized as $\Gamma_{\vb{k}+\vb{g},\vb{k}+\vb{g}'}$, where $\vb{k}$ belongs to the first Brillouin zone in reciprocal space and $\vb{g}$, $\vb{g}'$ are arbitrary vector in the reciprocal lattice $\mathcal{R}_{\vb{a}_i}$.
	
	\section{Lattice discretization}
	\label{app:discretization}
	
	Even after imposing the selection rules associated to the periodicity in a crystalline phase, the number of variational parameters (the Fourier coefficients $\psi_{\vb{k}}$ and $\Gamma_{\vb{k},\vb{k}'}$) is still infinite. In order to make the variational problem amenable to a numerical solution there are two possible approaches.
	The first is to truncate the Fourier expansion in order to include wave vectors with length smaller than a given threshold, namely $\abs{\vb{k}} < k_{\rm max}$. The second is to use a discrete lattice to approximate continuous space.
	The two approaches are similar since using a discrete lattice has also the effect of imposing an upper bound on wave vector magnitude. However, the second approach is simpler to implement in practice and it is the one used in the present work.
	
	The discrete numerical lattice used to approximate continuous space is defined by a pair of basis vectors $\vb{b}_i$. Then, the lattice site positions are labeled by a pair of integers $\vb{j}=(j_1,j_2)$ and are given by 
	\begin{equation}
		\label{eq:lattice_points}
		\vb{r}_{\vb{j}}  = j_1 \vb{b}_{1} + j_2\vb{b}_{2}.
	\end{equation}
	It is convenient to take the basis vectors $\vb{b}_i$ commensurate with the basis vectors $\vb{a}_i$ that specify the periodicity of the crystalline phase. The simplest way to impose commensurability is to require
	\begin{equation}
		\label{eq:a_Mb}
		\vb{a}_i = M_i \vb{b}_i\,,
	\end{equation} 
	where $M_i$ with $i = 1,2$ is a pair of positive integers, which should be taken as large as possible in order to well approximate the continuum problem. As a consequence of~\eqref{eq:R_rel_a}, one also has the relation $\vb{R}_i = M_iL_i\vb{b}_i$, moreover the total number of lattice sites of the numerical lattice contained in a supercell is $N_{\rm s} = \prod_{i = 1,2}M_iL_i$.
	
	After the discretization, the condensate wave function becomes a function of the pair of integers $\vb{j} = (j_1,j_2)$ labeling the lattice sites in the numerical lattice and the convention for the Fourier expansion is slightly modified
	\begin{equation}
		\label{eq:psi_Four_conv_1_disc}
		\psi(\vb{j}) = \frac{1}{\sqrt{N_{\rm s}}} \sum_{\vb{k}} \psi_{\vb{k}}e^{i\vb{k}\cdot\vb{r}_{\vb{j}}}\,.
	\end{equation}
	The number of wave vectors in the Fourier expansion is finite since two wave vectors that differ by a reciprocal lattice vector  are identified, namely
	\begin{equation}
		\label{eq:identification}
		\vb{k}\equiv \vb{k}' \Leftrightarrow \vb{k} -\vb{k}' \in \mathcal{R}_{\vb{b}_i}\,.
	\end{equation}
	Here, the reciprocal lattice $\mathcal{R}_{\vb{b}_i}$ associated to the basis vectors $\vb{b}_i$ is defined as in~\eqref{eq:rec_lattice_def}.  
	Moreover, the condensate wave function on the discrete lattice $\psi(\vb{j})$ is a dimensionless quantity since $\abs{\psi(\vb{j})}^2$ is the number of particles on site $\vb{j}$. Indeed, the mapping from the continuum to discrete space is performed according to the prescription
	\begin{equation}
		\sqrt{A_{\vb{b}_i}}\psi(\vb{r}_{\vb{j}}) \to \psi(\vb{j})\,,\quad
		\sqrt{A_{\vb{b}_i}}\hat{\psi}(\vb{r}_{\vb{j}}) \to \hat{\psi}(\vb{j})\,,
	\end{equation} 
	where $A_{\vb{b}_i} = \abs{\vb{b}_1 \times \vb{b}_2} = A/N_{\rm s}$ is the area of the unit cell of the lattice generated by the basis vectors $\vb{b}_i$.
	On the other hand, the Fourier coefficients $\psi_{\vb{k}}$ are dimensionless, both in~\eqref{eq:psi_Four_conv_2} and~\eqref{eq:psi_Four_conv_1_disc}.

	As an example of the discretization procedure, the condensate interaction energy appearing in the Gross Pitaevskii functional~\eqref{eq:GP_functional} becomes
	
	\begin{equation}
		\label{eq:example_discretization}
		\begin{split}
			&\frac{1}{2}\iint \dd{\vb{r}}\dd{\vb{r}'}\abs{\psi(\vb{r})}^2\abs{\psi(\vb{r}')}^2V(\vb{r}-\vb{r}')  \\ 
			&\hspace{1cm}\to \frac{1}{2} \sum_{\vb{j}}\sum_{\vb{j}'}\abs{\psi(\vb{j})}^2\abs{\psi(\vb{j}')}^2V(\vb{r}_{\vb{j}}-\vb{r}_{\vb{j}'}) \\
			&\hspace{1cm}=\frac{1}{2N_{\rm s}} \sum_{\vb{k},\vb{k}',\vb{q}} V_{\vb{q}}\psi_{\vb{k}}^*\psi_{\vb{k}-\vb{q}} \psi_{\vb{k}'}^*\psi_{\vb{k}'+\vb{q}}\,,
		\end{split}
	\end{equation} 
	where the Fourier coefficients of the interaction potential are defined as
	\begin{equation}
		\label{eq:Vq_discrete}
		V_{\vb{q}}	= \sum_{\vb{j}} e^{-i\vb{q}\cdot \vb{r}_{\vb{j}}}
		V(\vb{r}_{\vb{j}})\,, \quad V_{\vb{q}} = V_{\vb{q}}^* = V_{-\vb{q}}\,.
	\end{equation}
	The last sum in~\eqref{eq:example_discretization} is finite since it is performed over all wave vectors satisfying~\eqref{eq:kR_cond} with $0 \leq m_i < M_iL_i$. Indeed, one has to keep in mind that wave vectors are identified modulo reciprocal lattice vectors according to~\eqref{eq:identification}.
	
	Note that, in constrast to~\eqref{eq:Fourier_V}, the Fourier coefficients in~\eqref{eq:Vq_discrete} are obtained by sampling the interaction potential at the discrete points of the numerical lattice and thus depend on the discretization used (the choice of the vectors $\vb{b}_i$). To avoid discontinuous changes in the Fourier coefficients when the lattice vectors $\vb{b}_i$ are changed, the interaction potential is a slightly  smoothed version of the soft-core potential in~\eqref{eq:soft_core}
	\begin{equation}
		V(\vb{r}) = \frac{W}{e^{30(r-R)/R}+1}\,.
	\end{equation}
	The dependence of the Fourier coefficients on the discretization is one disadvantage of using the second approach mentioned at the beginning of the section. Its  consequences are discussed further in the following.
	
	In order to complete the mapping of the continuum Hamiltonian to a lattice Hamiltonian, it is necessary to specify the hopping matrix elements between the lattice sites. There is considerable freedom in how to choose them since the only constraint is that the effective mass in the lattice model (related to the coefficient of the quadratic term of the dispersion at $\vb{k}=0$) should be the same as the mass $m$ in the continuum. The specific choice used in this work is explained in the following.
	
\begin{figure}
		\includegraphics[scale=1]{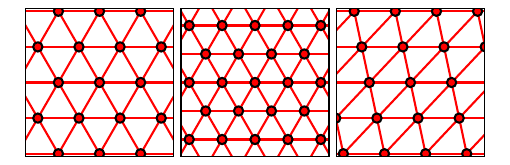}
		\caption{\label{fig:sheared_lattice} Left: Triangular lattice generated by the vectors $\vb{b}_i$~\eqref{eq:vec_triangular_sheared}  with $\theta = 2\pi/3$ or equivalently by $\vb{a}_i$~\eqref{eq:a_Mb}. In the former case, the lattice shown represents the numerical lattice used to map the continuum Hamiltonian to a discrete one, while in the latter it is a schematic representation of the spontaneously formed periodic structure that defines the solid phase. Middle: the same triangular lattice in the left panel rescaled by a factor $1-\lambda = 0.8$, where $\lambda$ parametrizes the hydrostatic compression used to computed the bulk modulus~\eqref{eq:hydro_comp}.
		Right: Lattice generated by the vectors $\vb{b}_i$~\eqref{eq:vec_triangular_sheared} with $\theta = 2\pi/3 - \pi/10$, which corresponds to a pure shear deformation of the triangular lattice in the left panel. From the change of the mean-field free energy $F_{\rm m.f.}$ with the angle $\theta$  one obtains the shear modulus according to~\eqref{eq:F_theta}.}
	\end{figure}

	In the case of the two-dimensional soft-core interaction potential, the most favorable crystalline symmetry is hexagonal~\cite{Pomeau1994}. For the purpose of computing the shear modulus (Sec.~\ref{sec:BKT_transitions}), it is necessary to consider a less symmetric numerical lattice generated by the vectors
	\begin{gather}
		\label{eq:vec_triangular_sheared}
		\vb{b}_1 = b\pmqty{1 \\ 0},\hspace{1.4mm}
		\vb{b}_2 = \frac{b\sqrt{3}}{2}\pmqty{\cot\theta \\ 1},\hspace{1.4mm}
		\vb{b}_3 = \vb{b}_1 + \vb{b}_2\,.
	\end{gather}
	For $\theta = \frac{2\pi}{3}$ and $\cot \frac{2\pi}{3} = -\frac{1}{\sqrt{3}}$, the triangular lattice with $C_6$ symmetry is recovered as a special case. For generic values of the angle $\theta$, the Bravais lattice is a monoclinic one, in which the rows of lattice sites are shifted horizontally so has to preserve the distance between them. This deformation preserves the area of the primitive unit cell and corresponds to a pure shear strain. The triangular lattice and its deformed version are illustrated in Fig.~\ref{fig:sheared_lattice}. 
	The only nonzero hopping amplitudes are the ones that correspond to nearest-neighbors in the triangular lattice with $\theta = 2\pi/3$. Thus, the dispersion is given by
	\begin{equation}
		\label{eq:disp_num_lat}
		\varepsilon_{\vb{k}} = \sum_{i=1}^32t_i(\theta)(1-\cos\vb{k}\cdot\vb{b}_i)\,. 
	\end{equation}
	The hopping amplitudes along the three directions specified by $\vb{b}_i$ are denoted by $t_i(\theta)$ and are given by
	\begin{gather}
		t_1(\theta) = t\qty[\frac{3}{2}\qty(1+\cot^2{\theta})+\sqrt{3}\cot\theta]\,,\\
		t_2(\theta) = 	t(2+\sqrt{3}\cot\theta) \,, \\
		t_3(\theta) = -t\sqrt{3}\cot\theta\,.
	\end{gather}
	The effective mass tensor at $\vb{k}=0$ is diagonal and independent of $\theta$. Moreover, if one fixes the hopping amplitude $t$ as
	\begin{equation}
		t = \frac{2R^2\epsilon_0}{3b^2} \qq{with} \epsilon_0 = \frac{\hbar^2}{2mR^2}\,, 	
	\end{equation}  
	the effective mass is equal to the mass $m$ in the continuum Hamiltonian, as required.

	\section{Iterative solution of the self-consistency equations}
	\label{app:iterative_solution}
	
	With the discretization procedure presented in Appendix~\ref{app:discretization}, the problem is reduced to the minimization of the mean-field grand potential $\Omega_{\rm m.f.}$ with respect to a finite number of variables $\psi_{\vb{k}}$, $\Gamma_{\vb{k},\vb{k}'}$, where the wave vectors $\vb{k}$, $\vb{k}'$ satisfy the condition~\eqref{eq:kR_cond} with $0 \leq m_i < M_iL_i$.  The number of independent variables is further reduced by imposing the periodicity conditions~\eqref{eq:psi_cond_per}-\eqref{eq:Gamma_cond_per}, which lead to the selection rules in~\eqref{eq:psi_sel_rule} and~\eqref{eq:f_sel_rule}. The special case of a translationally invariant system is obtained for $\vb{a}_i = \vb{b}_i$ ($M_i=1$), see~\eqref{eq:psi_c_kc}-\eqref{eq:Gamma_tr_inv}.
	
	In practice, rather then performing the simultaneous minimization of $\Omega_{\rm m.f.}$ with respect to all the independent variables, the strategy used in the present work is to seek a self-consistent solution by an iterative procedure. This is an adaptation to the bosonic case of the standard iterative method routinely adopted for fermionic systems. Unfortunately, the iterative algorithm to be presented below is found to be highly unstable when the pairing potential $\Delta(\vb{r},\vb{r}')$ is included as a variational parameter. This is the sole reason why the pairing potential is not used here. Finding a suitable algorithm for the  solution of the self-consistency equations of HFB theory, that is with the pairing potential included, is an interesting open problem for the future.

	The iterative algorithm for HF theory consists in the following steps:
	\begin{enumerate}
		\item Initialization: the energy functional~\eqref{eq:F_eGP_k_space} is minimized at fixed  density
		\begin{equation}
			\label{eq:condensate_rho}
			\rho = \frac{1}{A}\sum_{\vb{k}} \abs{\psi_{\vb{k}}}^2\,.
		\end{equation}
		The expectation values $\ev*{\hat{b}_{\vb{k}}^\dg\hat{b}_{\vb{k}'}}$ appearing in the functional are initially set to zero. The minimization produces an initial guess for the condensate wave function $\psi_{\vb{k}}$ and the chemical potential $\mu$, which is the Lagrange multiplier associated to the density constraint.
		
		\item Initialization: given the condensate wave function, the self-consistency equations~\eqref{eq:sc_cont_1_ksp}-\eqref{eq:sc_cont_3_ksp} with $\ev*{\hat{b}_{\vb{k}}^\dg\hat{b}_{\vb{k}'}}=0$ are used to provide an initial guess for  $\Gamma_{\vb{k},\vb{k}'}$.
		
		\item The quasiparticle Hamiltonian $H_0$
		 is constructed using the chemical potential and the HF potential obtained in the previous steps, see~\eqref{eq:H0_k_space_single}.
		
		\item From the quasiparticle Hamiltonian, the one-body density matrix at temperature $T$ ($\beta = 1/k_{\rm B}T$ is the inverse temperature) is computed as
		\begin{equation}
			\label{eq:one-body_density_mat}
			R = \frac{1}{e^{\beta H_0}-1}\,.
		\end{equation}
		The matrix elements of the density matrix $R$ are the expectation values $\ev*{\hat{b}_{\vb{k}}^\dg \hat{b}_{\vb{k}'}}$. 	 Note that at zero temperature the HF approximation reduces to the solution of the standard GP equation (step 1 above) since the one-body density matrix~\eqref{eq:one-body_density_mat} vanishes. 
		
		\item The energy functional~\eqref{eq:F_eGP_k_space}, in which the expectation values $\ev*{\hat{b}_{\vb{k}}^\dg \hat{b}_{\vb{k}'}}$ are the ones computed in the previous step, is minimized at fixed density
		\begin{equation}
			\label{eq:total_density}
			\rho = \frac{1}{A}\sum_{\vb{k}}\qty(\abs{\psi_{\vb{k}}}^2 + \ev*{\hat{b}_{\vb{k}}^\dg \hat{b}_{\vb{k}}})\,.	
		\end{equation}
		In contrast to~\eqref{eq:condensate_rho}, the density contribution of the quasiparticle excitations is now included in the total density.  
		The minimization produces updated values for the chemical potential $\mu$ and the condensate wave function $\psi_{\vb{k}}$.
		
		\item The condensate wave function $\psi_{\vb{k}}$ and the expectation values $\ev*{\hat{b}_{\vb{k}}^\dg \hat{b}_{\vb{k}'}}$ obtained respectively at steps 5 and 4 are used in the self-consistency equations~\eqref{eq:sc_cont_1_ksp}-\eqref{eq:sc_cont_3_ksp} to obtain an updated HF potential $\Gamma_{\vb{k},\vb{k}'}$.
		
		\item If the value of the grand potential $\Omega_{\rm m.f.}$, the condensate wave function and the HF potential have converged within a given threshold, the procedure terminates. Otherwise, the iteration continues from step 3.  
	\end{enumerate}
	If one seeks a self-consistent solution at fixed chemical potential rather than fixed total density $\rho$, the only modification is that the chemical potential term $-\mu \rho$, with the density given by~\eqref{eq:total_density}, must be added to the energy functional~\eqref{eq:F_eGP_k_space} and the minimization at steps 1 and 5 is performed in an unconstrained manner.

	Note that, according to the above sequence, the quasiparticle Hamiltonian $H_0$ computed at step 3 is in fact the upper diagonal block of the Hessian of the GP functional computed at its minimum (step 1 or 5), see~\eqref{eq:Hessian}-\eqref{eq:Hessian_2}. This is so by design and guarantees that the quasiparticle Hamiltonian $H_0$ is a positive definite operator. This means that the quasiparticle excitation energies, the eigenvalues of $H_0$, are positive, an important requirement for bosonic systems.
	
	The above iterative scheme is very stable and converges rapidly to a self-consistent solution in the same way as its fermionic counterpart. 
	The most computationally expensive step is the diagonalization of the quasiparticle Hamiltonian $H_0$, which is required to compute the one-body density matrix (step 4). By taking advantage of discrete translational symmetry, the diagonalization of $H_0$~\eqref{eq:H0_k_space_single} with size $\prod_i M_iL_i$, is reduced to the diagonalization of $L_1L_2$ blocks of size $M_1M_2$. It is found in practice that all observables of interest converge rather rapidly with the parameter $L_i = L$, specifying how finely the first Brillouin zone is sampled. The value $L=6$ is found to provide very well converged results and is always used here. On the other hand, the convergence is found to be substantially slower with respect to the parameter $M_i=M$, which is much harder to increase given that the computational cost scales as $M^6$. A qualitatively good phase diagram can be obtained for $M=14$, however we have found it necessary to increase this discretization parameter further to ensure convergence. For the results shown here the value $M=22$ is generally used, which is at the edge of our computational capabilities.
	
	\begin{figure}
	\includegraphics{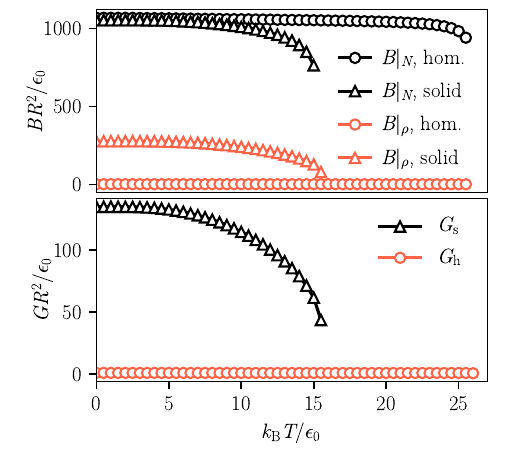}
	\caption{\label{fig:bulk_shear_moduli} Upper panel: bulk modulus in the homogeneous and solid phases as a function of temperature. The bulk modulus is computed by rescaling the supercell, which is equivalent to a hydrostatic compression, see~\eqref{eq:hydro_comp} and Fig.~\ref{fig:sheared_lattice}, while keeping either the total number of particle ($\left.B\right|_{N}$) or the average density ($\left.B\right|_\rho$) constant. In the homogeneous phase $\left.B\right|_\rho$ should vanish, however, due to the discretization procedure used in this work (Appendix~\ref{app:discretization}), it is not exactly zero, but nevertheless very small, $B|_{\rho} \lesssim 1.6 \epsilon_0/R^2$. In the solid phase, $\left.B\right|_\rho$ is nonzero since a rescaling of the supercell also changes the lattice constant $a=\abs{\vb{a}_i}$, see~\eqref{eq:R_rel_a}. Bottom panel: shear modulus in the homogeneous ($G_{\rm h}$) and solid ($G_{\rm s}$) phases as a function of temperature. Due to the mapping to a discrete lattice model, the shear modulus in the homogenenous phase $G_{\rm h}$ is not exactly zero, as expected for a fluid. However, for large enough discretization parameter, here $M=22$, it is essentially negligible, $G_{\rm h} \lesssim 0.8\epsilon_0/R^2$. All parameters are the same as in Fig.~\ref{fig:three_panels}.}	
	\end{figure}

	\begin{figure*}
		\includegraphics{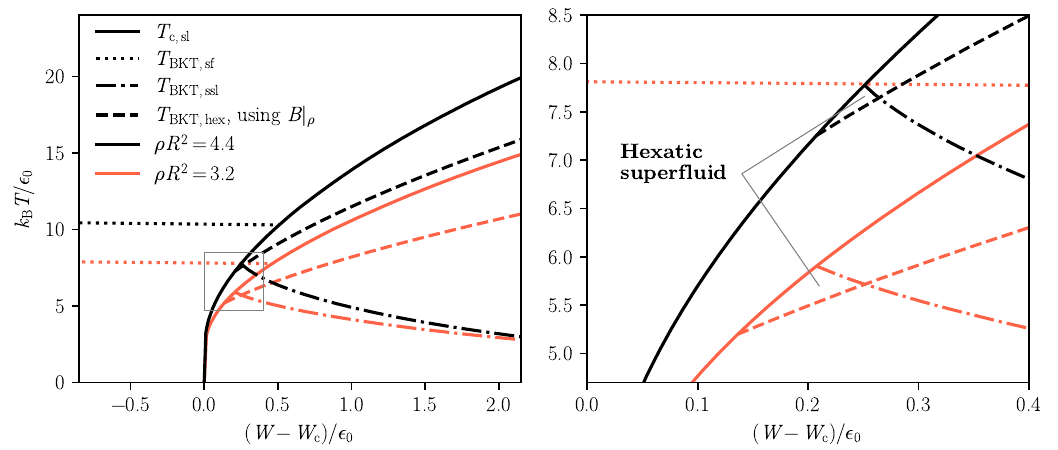}
		\caption{\label{fig:pd_alternative} Left: phase diagram of soft-core bosons as in Fig.~\ref{fig:phase_diagram}, with the only difference that the BKT transition temperature  from the solid to the hexatic phase $T_{\rm BKT, hex}$ is obtained from~\eqref{eq:KN_relation_melting}-\eqref{eq:Young}, in which the bulk modulus $B$ is computed at constant density $\rho$, denoted as $B|_{\rho}$ in Fig.~\ref{fig:bulk_shear_moduli}, rather than at constant particle number, denoted as $B|_{N}$. As a result, the phase boundary (dashed line) is at lower temperature compared to Fig.~\ref{fig:phase_diagram} and there is a small overlap between the supersolid and hexatic regions, indicating the presence of an hexatic superfluid phase. Right: magnified view of the phase diagram corresponding to the square on the left panel. The small hexatic superfluid region becomes larger with decreasing density.  The coupling constant $W$ on the horizontal axis is shifted by $W_{\rm c}$, which is the critical value above which a spontaneous modulation of the particle density is energetically favoured, see Fig.~\ref{fig:lattice_constant}. The shift of the coupling constant facilitates the comparison of the phase diagrams for different particle densities. The critical values are $W_{\rm c} = 5.85\epsilon_0$ at $\rho R^2 = 4.4$ and $W_{\rm c} = 8.04\epsilon_0$ at $\rho R^2 = 3.2$.} 
	\end{figure*}
	
	\section{Bulk and shear moduli}
	\label{app:bulk_shear}
	
	The bulk and shear moduli as a function of temperature at fixed coupling constant $W=7\epsilon_0$ and density $\rho R^2 =4.4$ are shown in Fig.~\ref{fig:bulk_shear_moduli}. The bulk modulus is obtained from the change of the mean-field free energy under hydrostatic compression according to~\eqref{eq:B_rho_def}. A pure hydrostatic compression is implemented in the numerical HF calculations by rescaling the fundamental vectors $\vb{R}_i$, $\vb{a}_i$ and $\vb{b}_i$ by the same factor $(1-\lambda)$, where  the coefficient $\lambda$ appears in the displacement field~\eqref{eq:hydro_comp}. These are vectors that specify the size of the supercell, the periodicity of the crystalline structure and the fineness of the discretization, respectively, and are all proportional to each other according to~\eqref{eq:R_rel_a} and~\eqref{eq:a_Mb}. The bulk modulus, which is used to obtain the Young modulus and the BKT temperature $T_{\rm BKT, hex}$ shown in Fig.~\ref{fig:three_panels}c, measures the change of the free energy density under compression for fixed particle number. It is indicated with the symbol $B|_{N}$ in Fig.~\ref{fig:bulk_shear_moduli} and is the inverse of the thermodynamic isothermal compressibility. 
	
	However, it is also possible to compute the change of the free energy density under compression at fixed average density $\rho$, leading to an alternative definition of bulk modulus, which is denoted as $B|_{\rho}$ and shown in Fig.~\ref{fig:bulk_shear_moduli} as well. Note that in the homogeneous phase $B|_{\rho}$ is zero since the free energy density is purely a function of the density and the temperature, which are constant. Due to the mapping of the continuum Hamiltonian to a discrete lattice model, $B|_{\rho}$ is nonzero in the homogeneous phase, but essentially negligible  for large enough discretization parameter $M$. The reason is that the Fourier transform of the interaction potential~\eqref{eq:Vq_discrete} depends on the vectors $\vb{b}_i$, see~\eqref{eq:lattice_points}. On the other hand, in the solid phase the free energy density is a function of an additional parameter, namely the lattice constant $a$ of the triangular lattice formed by the spontaneous density modulations, thus $B|_{\rm \rho}$ is nonzero since it  measures the energy cost associated to the change of the lattice constant alone. 
	
	In the theory of elasticity, the distinction between $B|_{\rho}$ and $B|_{N}$ is not made since for a standard crystalline material the electronic density and the lattice constant are not independent of each other. On the other hand, for a cluster solid of the type formed by soft-core bosons, the number of particles per unit cell is not strictly an integer and, for fixed $\rho$, the solid phase can be favored with respect to the homogeneous one within a finite interval of values for the lattice constant. Thus, for a finite size system the lattice constant is controlled by the boundary conditions and it becomes necessary to determine the optimal lattice constant by requiring the vanishing of the first derivative of the free energy density with respect to $\lambda$ at constant density $\rho$. This is the prescription used to compute $a_{\rm opt}$ shown in the upper panel of Fig.~\ref{fig:lattice_constant}. The optimal lattice constant is the preferred periodicity of the density modulations realized in an infinite system, which is not constrained by the boundary conditions.

	The distinction between the two possible definitions of bulk modulus is necessarily important for a supersolid since superfluid and crystalline orders can not be present simultaneously if the number of particles is commensurate with the lattice structure, that is if the number of particles per unit cell is an integer \cite{Prokofev2005}. Note that similar issues regarding the bulk modulus in the case of the soft-core boson model have been discussed also in a recent work~\cite{Rakic2024}. On the other hand, this problem has not been considered so far in relation to the theory of two-dimensional melting, meaning that is not clear whether $B|_{N}$ or $B|_{\rho}$ should be used in~\eqref{eq:Young} for computing the Young modulus and  estimating the BKT temperature $T_{\rm BKT,hex}$. 
	
	For completeness, Fig.~\ref{fig:pd_alternative} shows the same phase diagram as in Fig.~\ref{fig:phase_diagram}, with the only difference that the BKT temperature $T_{\rm BKT,hex}$ is estimated from~\eqref{eq:KN_relation_melting} and~\eqref{eq:Young} using $B|_{\rho}$ instead of $B|_{N}$. Moreover, the phase diagrams for two different values of the density are compared ($\rho R^2 = 3.2,\,4.4$). As seen from Fig.~\ref{fig:bulk_shear_moduli}, $B|_{\rho}$ is roughly a factor four smaller than $B|_{N}$, therefore the transition line between solid and hexatic phases is shifted to lower temperatures in Fig.~\ref{fig:pd_alternative} compared to Fig.~\ref{fig:phase_diagram}, leading to a small overlap between the hexatic and supersolid regions, namely an hexatic superfluid phase~\cite{Mullen1994}.
	
	\section{Finite size effects in PIMC simulations}
	\label{app:finite_size_pimc}
    \begin{figure*}
        \includegraphics{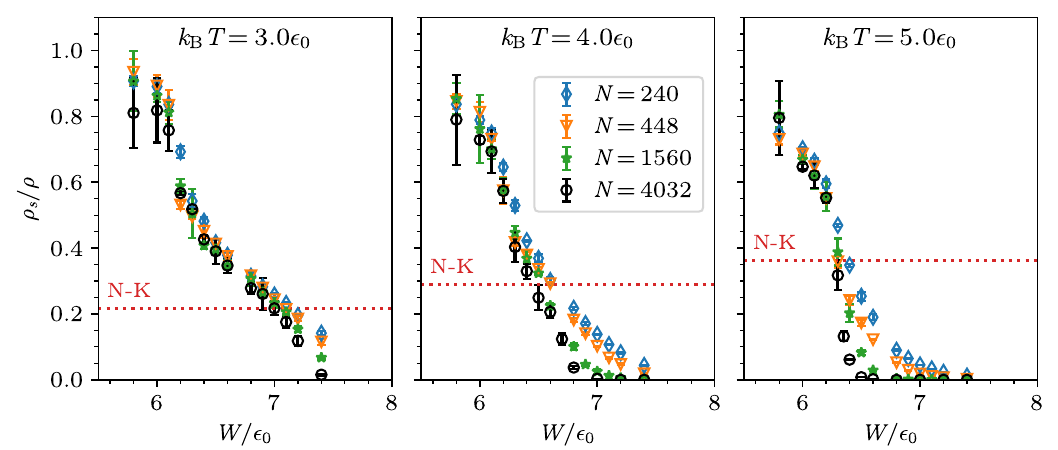}\\
        \includegraphics{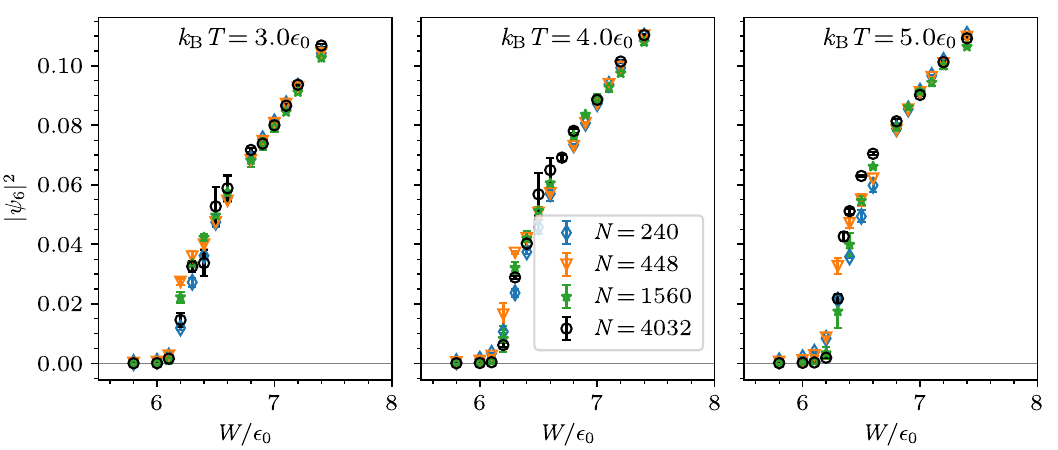}
        \caption{\label{fig:pimc_scaling}Finite size scaling of the order parameters in PIMC simulations. Upper row: superfluid fractions computed at four system sizes $N=240,\,448,\,1560,\,4032$. The finite size effects are sizable at stronger values of the interaction, after the superfluid fraction crosses the Nelson-Kosterlitz prediction (red dotted line). Lower row: the orientational order parameter for the same system sizes shows much smaller dependence on the system size.}
        \label{fig:pimc_sf_scaling}
    \end{figure*}
    The phase diagram shown in the right panel of Fig.~\ref{fig:phase_diagram} is obtained from PIMC simulations with $N=4032$ particles. Additional sizes are considered, including the larger size $N=16128$ (see~Fig.~\ref{fig:pimc_correlations}), as well as  smaller sizes, as discussed hereafter. This allows us to verify when  finite size effects occur. 
    In Fig.~\ref{fig:pimc_scaling} we show the superfluid fraction and the orientational order parameter at three temperatures, for various system sizes. The black points at $N=4032$ are the same data reported in Fig.~\ref{fig:pimc_order_param}. It is worth reminding that, for increasing interaction strength $W$, the system transitions from the superfluid to the solid phase, passing through a supersolid phase in the cases corresponding to the two lowest temperatures $k_\mathrm{B}T = 3.0\epsilon_0$ and $4.0\epsilon_0$.
    Instead, at $k_\mathrm{B}T = 5.0\epsilon_0$ a direct transition from superfluid to normal solid is obtained. 
    As shown in the upper row of Fig.~\ref{fig:pimc_scaling}, the finite size effects on the superfluid fraction are very modest in the weak interaction regime, as long as the system is superfluid or supersolid.
    However, as the superfluid fraction crosses the Nelson-Kosterlitz prediction, and the system is hence classified as a normal solid, the finite size effects become important. This is consistent with the expectation of a BKT transition renormalizing the superfluid fraction to zero from the point predicted by the Nelson-Kosterlitz relation. On the other hand, the global orientational order $|\psi_6|^2$ (see second row of  Fig.~\ref{fig:pimc_scaling}) shows little dependence on the system size. The only exceptions are the points very close to the superfluid to (super)solid transition, which display large statistical fluctuations. In fact, for these interactions strengths, the phase is classified as unknown according to the analysis of the correlation functions discussed in Section~\ref{sec:PIMC}.

\input{phase_diagram_soft_disks.bbl}

\end{document}

%% file: phase_diagram_soft_disks.bbl
%